\pdfoutput=1
\documentclass[%
 reprint,
 amsmath,amssymb,
 aps,
pra,
floatfix,
]{revtex4-1}

\usepackage{graphicx}
\usepackage{dcolumn}
\usepackage{bm}
\usepackage{subfigure}
\usepackage{hyperref}
\usepackage{wrapfig}
\usepackage{placeins}
\usepackage{natbib}
\usepackage{morefloats}
\usepackage{bbold}
\usepackage{xcolor}

\begin{document}


\title{Influence of generic quantum coins on the spreading and entanglement in binary aperiodic quantum walks 
}
\author{Tushar Kanti Bose}
\email{tkb.tkbose@gmail.com}
\affiliation{Department of Physics, Srikrishna College, Bagula, Nadia, West Bengal-741502, India.}

\date{\today}
\begin{abstract}
Exploring the quantum walk as a tool of generating various probability distributions and quantum
entanglements is a topic of current interest. In the present work, we use extensive numerical simulations to investigate the influence of 
generic quantum coins on the hybrid entanglement and spreading behavior of different binary quantum walks with 
time and position dependent coin operations based on the Fibonacci, Thue-Morse and Rudin-Shapiro sequences. We find that each considered 
walk is differently but significantly influenced by the choice of quantum coins. We demonstrate that
the dynamic Fibonacci walk exhibits localized behavior for certain coin parameters. 
This behavior brings new information about the role played by dynamic coin disorder, 
considered before as always a generator of non-localized behavior. We also reveal
the significant role played by a specific coin parameter which controls the nature of superposition 
of spin up and spin down states during coin operation. We find that the parameter can distinctly tune the 
spreading and entanglement behavior of a binary quantum walk. We show that an increase in the value of the parameter can enhance both the the standard deviation of the position distribution of the walker and the hybrid entanglement from
significant low to significantly high values depending on the coin and the nature of coin operations. The present work may thus be considered
as one step towards understanding the role of coins in
inhomogeneous quantum walks.

\end{abstract}

\pacs{Valid PACS appear here}
\maketitle

\section{Introduction}

Quantum random walk or quantum walk (QW) in its original form is simply the dynamics of a quantum particle that has a spin-1/2-like internal degree of freedom in addition to its position and momentum \cite{first}. Being a natural quantum version of the classical random walk that appears in statistics, computer science, finance, physics, chemistry and biology, it has been a topic of fundamental interest \cite{review}. Moreover, QW research now enjoy broader interest due to its widespread applications in the areas of quantum algorithms \cite{kempe}, quantum computing \cite{q_computing}, quantum biology \cite{q_biology} and quantum simulation \cite{q_simulation}.\\ 

In the generic discrete time QW, the dynamics of a quantum walker is usually controlled by two unitary operators : a rotation operator \(\hat{C}\) (called ``quantum coin") and a shift operator \(\hat{S}\). The coin operator acts on the walker's internal degrees of freedom, leaving it generally in a superposition of spin up and spin down. The shift operator then shifts the position according to the walker's internal degree of freedom. Hence, the internal and external degrees of freedom becomes entangled. Successive applications of the two operators (\(\hat{C}\) \& \(\hat{S}\)) generate discrete time evolution of the walker. This is what we call the one-dimensional discrete time QW. Discrete time QWs can be implemented using NMR \cite{nmr}, ion traps \cite{ion}, waveguide arrays \cite{waveguide1,waveguide2} and superconducting devices \cite{superconducting}. One major advantage of QW over the classical random walk is that a quantum walker spreads over the line linearly in time (standard deviation \(\sigma\sim t\)), while the classical random walk spreads in a slower fashion (\(\sigma\sim t^{1/2}\)). Another crucial property of QW is the quantum entanglement between the coin and positional degrees of freedom. Quantum entanglement is of utmost importance for research in the areas of quantum information and quantum communications \cite{qq23,qq24,qq25,qq26,qq27}. The generation of quantum entanglement has also been realized in different experminents on QWs \cite{entanglement1,entanglement2}. \\


Disorders are ubiquitous in physical systems and thus it is natural that the study of discrete time QW subjected to different types of disorder/inhomogeneity has been drawing attention of several authors \cite{q25,q26,q27,q28,q29,q30,q31,q32,q33,q34,q35,q36,q37,q38,q39,q40,q41,q42,q43,q44,q45,q46,q47,q48,q49}. 
 In general, two different categories of disorder are considered : Dynamic (time-dependent) and Static (time-independent). One way to generate dynamic disorder is to use a randomly chosen coin at each time step 
\cite{dynamic1,dynamic2}. On the other hand, static disorder can be generated by placing a randomly chosen coin at each lattice point. These coins remain unchanged during the time evolution of the walker \cite{static1,static2,static3}. Besides considering coin-disorder, another way to generate dynamic/static disorder is to consider time/position dependent phase defects \cite{phasedefect1,phasedefect2,phasedefect3,phasedefect4}. Disorder can also be introduced in step lengths of the walker \cite{q46,step1}. Here we focus on coin disorder.

It was found that the coin disorder/inhomogeneity induces drastic deviations from the quadratic spreading behavior. For dynamic random disorder, there is diffusive spreading(\(\sigma \sim t^{1/2}\)) of the walker \cite{aperiodic} whereas for static random disorder, the walker gets localized \cite{q31,disorder1}. Both the dynamic and static random disorder have also been studied experimentally \cite{entanglement2,disorder1}. It describes the relevance of such disordered systems in areas of quantum information science. Apart from random disorder, QWs with deterministic disorder have also naturally attracted significant attention as the physical systems with deterministic aperiodicity have long been a subject of interest since the discovery of quasicrystals in 1951 \cite{quasi}. Deterministically aperiodic physical systems are ordered without any translational invariance and thus can be regarded as intermediate between ordered and random disordered systems. 
A simple way to generate static aperiodic arrangement of coins is to consider only the central coin to be different from other coins. Such arrangement shows localization of the walker \cite{one1,one2}. A binary aperiodic QW was first studied by Rebeiro et al. \cite{aperiodic}. They showed that time-dependent arrangement of two coins according to binary Fibonacci sequence exhibits a sub-ballistic spreading characterized by sequence dependant slopes. Similarly, a sequence of two coins arranged according to Levy waiting time distribution also exhibited sub-ballistic spreading \cite{levy1}. Two coins distributed on the lattice following fractal Cantor sequence exhibited a more complex dynamics : sub-ballistic spreading upto certain time and then quite slower evolution of standard deviation \cite{cantor}. More recently, spreading behavior of binary QWs with time and position dependent coin operations based on the Fibonacci sequence, the Rudin Shapiro sequence and the Thue-Morse sequence were studied in Ref.\cite{chandrasekhar}. In all the three cases, static aperiodicity generated a slower spreading in comparison to the dynamic aperiodicity. Fibonacci and Rudin-Shapiro walks exhibited fastest and slowest spreadings respectively among the three sequences under both type of aperiodicity \cite{chandrasekhar}. Localization behavior was found only in case of static Rudin-Shapiro walk \cite{chandrasekhar}. An experimental realization of binary walk with position dependent coins was performed by Xue et al. in  \cite{ben13}. Instead of using only two different coins as considered in binary aperiodic QW, higher number of coins can also be used to realize aperiodicity such that the number of different coins increases with time \cite{time1,time2}. Such walks also exhibited various kinds of spreadings. Buarque and Dias studied a QW with position dependent coin operations based on a deterministically aperiodic sequence of multiple coins where the aperiodicity can be tuned using a single parameter \cite{Buarque}. Localization or delocalization were reported depending on the degree of aperiodicity. Here we focus only on binary aperiodic walks.  

Most of the earlier works on coin disordered quantum walks remained focussed mainly on the transport properties of the walks. However, recently, attention has been shared by the coin-position entanglement properties. 
One of the reasons is a surprising result 
that the systems with dynamical random disorder exhibit maximal coin-position entanglement, independent of the initial state of the walker \cite{maximal}. This result was in contrast to the conventional idea that introduction of disorder makes the system more classical. The phenomenon was also verified in an experiment \cite{entanglement2}. Probably, this interesting phenomenon motivated some of the more recent works on QW which were focussed on discovering an ideal sequence of coins(dynamic disorder) which will provide both ballistic diffusion and maximal entanglement together as such phenomenon will be highly beneficial for quantum communication purposes \cite{optimal,optimal2}. Static random disorder, on the contrary, has been found to be worse than the ordered QW in respect of quantum entanglement \cite{q45,q47}. 
There have only been a couple of studies on the entanglement properties of deterministically aperiodic QWs.  Panda et al. studied the aperiodic QWs following the Parrondo sequences \cite{panda}. Highly entangled states including maximally entangled states were generated independent of the phase of the initial state. Buarque and Dias found that unlike static random disorder, static aperiodicity can enhance entropy \cite{Buarque}. Liu et al. studied the entanglement generation capability of dynamic and static Fibonacci and Thue-Morse QWs \cite{aperiodic_entanglement}. They studied the entanglement generation for different parts of the infinite aperiodic sequences and reported the average behavior. They found that the entanglement of Fibonacci QW can reach the maximal value for dynamic aperiodicity whereas the entanglement of Thue-Morse QW can be close to the maximal value. For the static aperiodic QW, the entanglement of Fibonacci QW was also found to reach the maximal value which is quite interesting as static disordered QWs have failed to generate maximal entanglement. They concluded that the entanglement of static aperiodic QWs are greater than those of QW with static random disorder and homogeneous QWs. The classification of aperiodic systems according to the entanglement of QW was found to be different from that by the quantum phase transition of the aperiodic quantum Ising chain \cite{Quantum_ising} and that by degrees of sequence of disorder of those sequences \cite{degree_sequence}. 
\\


It is easy to understand that the quantum state evolution in binary QWs depend also on the character of the quantum coins. So far binary quantum walks have mainly been studied using generalized Hadamard coins. It is thus natural to ask how would coins from a more generic set influence the evolution of binary QWs. Here we study such influence on deterministically aperiodic binary QWs. Another source of interest behind the present work has been to reveal the influence of coin parameters on the hybrid entanglement for the first time in case of any binary QW. In difference to the the study by Liu et al. \cite{aperiodic_entanglement} in which spin-position entanglement had been averaged over different parts of the aperiodic sequences, we study the entanglement dynamics for a long aperiodic sequence whose length is larger than both the chain length and the number of quantum walk steps. \\
Another point of interest behind the present work has been the findings reported by Jing et al. in their study on the effect of generic quantum coins on the spreading behavior of the binary QWs with random disorder \cite{jing}. They found that sub-ballistic spreading could be found in binary dynamic disordered quantum walk if the two quantum coins satisfy the following condition :  \( e^{i(\theta_{1}-\phi_{1})}=e^{i(\theta_{2} -\phi_{2})}  \) where \(\theta_{1},\phi_{1},\theta_{2},\phi_{2}\) are coin parameters. Similarly, the same condition was found to be required for binary static disordered walk to exhibit sub-diffusive, diffusive and sub-ballistic spreadings. It is thus natural to ask whether any such relation exists for deterministically disordered QWs and whether such relations exist for hybrid entanglement ? Here we look for the answer of these questions. Moreover, here we also intend to find the role of the coin parameter which controls the nature of superposition of coin states, on the spreading and entanglement. With a motivation of finding answers to all the above questions here we use extensive numerical simulations to investigate influences of generic quantum coins on the hybrid entanglement and spreading behavior of binary QWs with time and position dependent coin operations based on the Fibonacci sequence, the Thue-Morse sequence and the Rudin-Shapiro sequence.\\


A thorough identification of the properties of any aperiodic QW based on coin parameter variation may help to use it as a generator of various probability distributions and hybrid entanglements. The present work thus may be considered as one step towards understanding the role of coins in such inhomogeneous quantum walks.\\

The paper is organized as follows. In the section \ref{two}, we describe the formalisms of binary aperiodic QWs. All the numerical results of our study are presented in section \ref{three}. Sub-sections \ref{d1}, \ref{d2} and \ref{d3} describe dynamic versions of Fibonacci, Rudin-Shapiro and Thue-Morse QWs respectively. Sub-sections \ref{s1}, \ref{s2} and \ref{s3} describe static versions of Fibonacci, Thue-Morse and Rudin-Shapiro QWs respectively. Sub-section \ref{s4} presents a discussion on maximal entanglement generation. In section \ref{four}, we draw the conclusions and present future pathways.\\

\section{Model \& Formalism \label{two}}

The relevant degrees of freedom for a single particle discrete-time QW on a line are the particle’s position \(x\) (with \(x \in z\)) on the line, as well as its coin state. The total Hilbert space is given by \(H_{Total} \in H_{P}\otimes  H_{C}\) , where \(H_{P}\) is spanned by the orthonormal position vectors \(\{|x\rangle\}\) and \(H_{C}\) is the two-dimensional coin space spanned by two orthonormal vectors which we denote as \(|\uparrow\rangle\) and \(|\downarrow\rangle\). Each step of the QW consists of two subsequent operations: the coin operation and the shift-operation. The coin operation, given by \(\hat{C}\), and acting only on \(H_{C}\) , allows for superpositions of different alternatives, leading to different moves. This operation is the quantum equivalent of randomly choosing which way the particle will move in case of classical random walk. Then, the shift operation \(\hat{S}\) moves the particle according to the current coin state, transferring this way the quantum superposition to the total state in \(H_{Total}\). The evolution of the system at each step of the walk can then be described by the total unitary operator.\\
\begin{equation}
\hat{U}\equiv \hat{S}(\hat{I}\otimes\hat{C})  
\end{equation}
where \(\hat{I}\) is the identity operator acting on \({H_{P}}\). A popular choice for \(\hat{C}\) is the Hadamard operator \(\hat{C}_{H}\):
\begin{center}
\begin{equation}
\hat{C}_{H}=\frac{1}{\sqrt{2}} \begin{pmatrix}
       1 & 1 \\[0.3em]
       1 & -1 
     \end{pmatrix}
\end{equation}
\end{center}
The shift operator is given by
\begin{equation}
\hat{S} = ( \sum\limits_{x} |x+1\rangle \langle x|)\otimes |\uparrow\rangle\langle\uparrow| + ( \sum\limits_{x} |x-1\rangle \langle x|)\otimes |\downarrow\rangle\langle\downarrow|  \end{equation}

In general \(C(x,t)\) is a two-dimensional unitary matrix,  \begin{equation}
\hat{C}=\left(\begin{array}{cc}
\sqrt{\rho} & \sqrt{1-\rho} e^{i\theta}\\
\sqrt{1-\rho} e^{i\phi} & - \sqrt{\rho} e^{i(\theta + \phi )}\end{array}\right).\label{Cgral}\end{equation} where \(0\leq \rho \leq 1\), \(0 \leq \theta \) and \(\phi \leq 2\pi\). For \(\rho = 0.5\), the coin generates an equal superposition of spin up and spin down states while acting on either one of the two. For \(\rho \neq 0.5 \), it creates an unequal superposition. \(\theta\) and \(\phi\) together controls the relative phase between up and down states in superposition. For \(\rho = 0\) and \(\rho = 1\) the walker exhibits oscillatory and uniform motion respectively. Those are the trivial cases whereas non-triviliaty appears for \(0 < \rho < 1\). For a homogeneous QW, the quantum coins \(C(x,t)\) are independent of both position and time. 
For the binary QW with dynamical deterministic aperiodicity, at each step the quantum coin is either \(C_{1}(\rho_{1},\theta_{1},\phi_{1})\) or \(C_{2}(\rho_{2},\theta_{2},\phi_{2}\))  according to a particular deterministic binary sequence. On the other hand, for the binary QW with static deterministic aperiodicity, at each lattice site \(x\), the quantum coin is the quantum coin is either \(C_{1}(\rho_{1},\theta_{1},\phi_{1})\) or \(C_{2}(\rho_{2},\theta_{2},\phi_{2}\))  according to a particular deterministic binary sequence and remains unchanged for all time steps. The quantum coins \(C_{1}(\rho_{1},\theta_{1},\phi_{1})\) and \(C_{2}(\rho_{2},\theta_{2},\phi_{2}\))  are the matrices described by Eq.(4) and \(\rho_{1},\theta_{1},\phi_{1},\rho_{2},\theta_{2},\phi_{2}\)  are time-independent and position independent parameters in both the cases of dynamic and static deterministic aperiodicity. 
We considered the following three popular aperiodic sequences.\\
 (1) The Fibonacci sequence : This is a well known aperiodic sequence. It is obtained by iteration of the recursive rule \(S_{n+1}=S_{n}S_{n-1}\) with \(S_{0}=0\) and \(S_{1}=01\). The first few sequences are as follows : 0, 0 1, 0 1 0, 0 1 0 0 1, 0 1 0 0 1 0 1 0.... The infinite sequence is as follows :  0 1 0 0 1 0 1 0 0 1 0 0 1 0 1 0 0 1 0 1 0 0 1 0 0..... \\
(2) The Rudin-Shapiro sequence : This sequence is defined by the following relation : \(S_{n}=(-1)^{n}\) where \(u_{n}=\sum_{k\geq 0}\epsilon_{k}\epsilon_{k+1}\). Here \(\epsilon_{k}\) is the coefficient obtained from the related binary conversion as follows : \(n=\sum_{k\geq 0} \:\epsilon_{k}\: 2^{k}\). 
The infinite sequence is as follows 1 1 1 -1 1 1 -1 1 1 1 1 -1 -1 -1 1 -1.......\\
(3) The Thue-Morse sequence : The Thue-Morse sequence is obtained by iteration of the recursive rule \(S_{N+1}=S_{N}\bar{S_{N}}\) with \(S_{0}=0\). Here \(\bar{S_{N}}\) is the string \(S_{N}\) with 0 and 1 replacing each-other. The infinite sequence is as follows 0 1 1 0 1 0 0 1 1 0 0 1 0 1 1 0 1 0 0 1 0 1 1 0 0....\\

We are not considering periodically repeated small approximants. We numerically generate a long sequence whose length is larger than both the chain length and the number of random walk steps.\\


We perform the simulations for five thousand random walk steps starting from the following initial state : \(  |\psi_{0}\rangle = \frac{1}{\sqrt{2}} (|0, \uparrow\rangle + i|0, \downarrow\rangle )\). The standard deviations of the position distributions of the walker are then calculated numerically for all time steps. We find that the standard deviation is dependent on time with \(\sigma \sim t^{\alpha}\). The value of the exponent \(\alpha\) is then extracted by fitting the data obtained from the last 4500 time steps to the above relation using the least square fit method. 
 We also numerically calculate the entanglement between the internal and external degrees of freedom using the Von Neumann entropy of the partially reduced state $\rho_C(t)=Tr_P(\rho(t))$ \cite{Bennett}, where $\rho(t)=|\Psi(t)\rangle\langle\Psi(t)|$ is pure and $Tr_P(\cdot)$ indicates trace over the
position degrees of freedom. Considering \( |\Psi(t)\rangle = \sum\limits_{j} ( a(j,t) |\!\uparrow \rangle |j\rangle + b(j,t)|\!\downarrow\rangle |j\rangle)\), we get  \( \rho_C(t) \!\!=\!\! \alpha(t) |\!\uparrow\rangle\langle\uparrow\!|\!+\!
\beta(t) |\!\downarrow\rangle\langle\downarrow\!| \!+\!
\gamma(t)\) 
$ |\!\uparrow\rangle\langle\downarrow\!| 
\!+\! \gamma^*(t) |\!\downarrow\rangle\langle\uparrow\!|,
$ where \( |\Psi(t)\rangle = \sum\limits_{j} ( a(j,t) |\uparrow \rangle |j\rangle + b(j,t)|\downarrow\rangle |j\rangle)\) and \(\alpha(t)=\sum_{j} |a(j,t)|^2, \beta(t)=\sum_{j} |b(j,t)|^2, \gamma(t)=\sum_{j} a(j,t)b^*(j,t)\). The entanglement is then given by \( S_E(\rho(t))=-Tr(\rho_C(t)\log_2\rho_C(t)) = -\lambda_+(t)log_2\lambda_{+}(t)-\lambda_{-}(t)log_2\lambda_{-}(t)\) with \( \lambda_{\pm}=(1/2{\pm}\sqrt{1/4-\alpha(t)(1-\alpha(t))+|\gamma(t)|^2})\) being the eigenvalues of $\rho_C(t)$. $S_E$ is $0$ for separable states and $1$ for maximally entangled ones.  \(S_E\) here is not only a function of time but also a function of the coin parameters \(\rho_1,\theta_1,\phi_1,\rho_2,\theta_2,\phi_2\). We calculate average entropy \(\langle S_{E} \rangle \) by performing an average over \(S_{E}\) values obtained for last 2500 steps. The corresponding error bar is obtained by calculating the standard deviation of the obtained \(S_{E}\) values over time. \\

\begin{figure*}[th]
\centering

\subfigure[\label{fig:1a} ]{\includegraphics[scale=0.40]{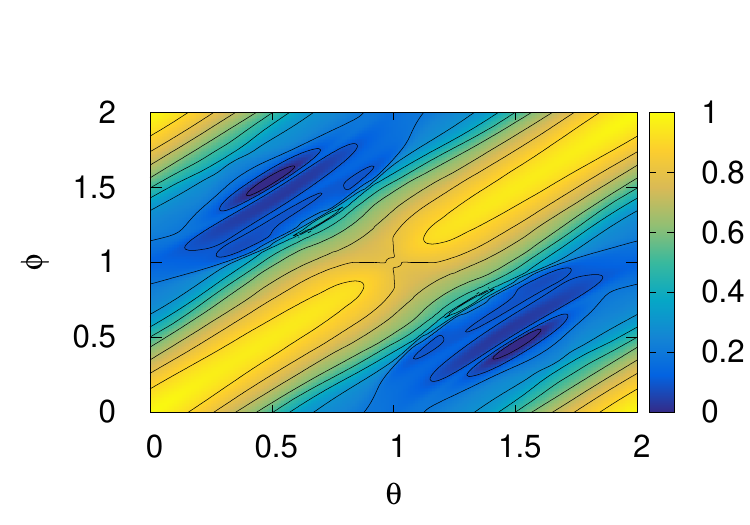}}\hspace*{.185cm}
\subfigure[\label{fig:1b} ]{\includegraphics[scale=0.40]{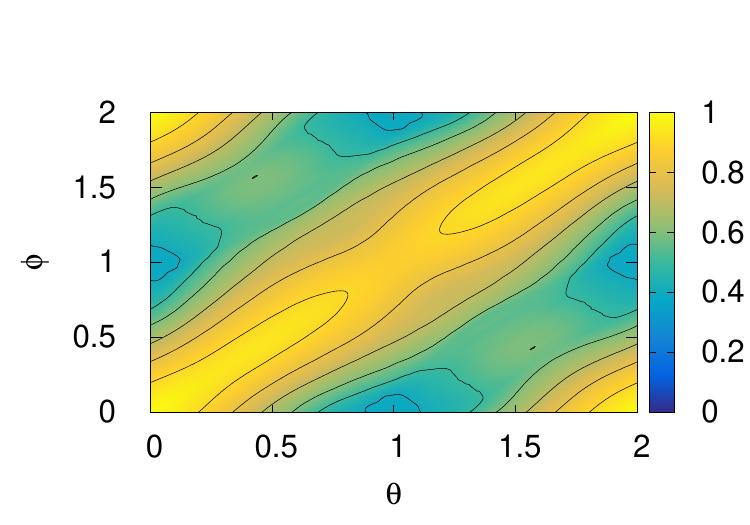}}\hspace*{.185cm}
\subfigure[\label{fig:1c} ]{\includegraphics[scale=0.40]{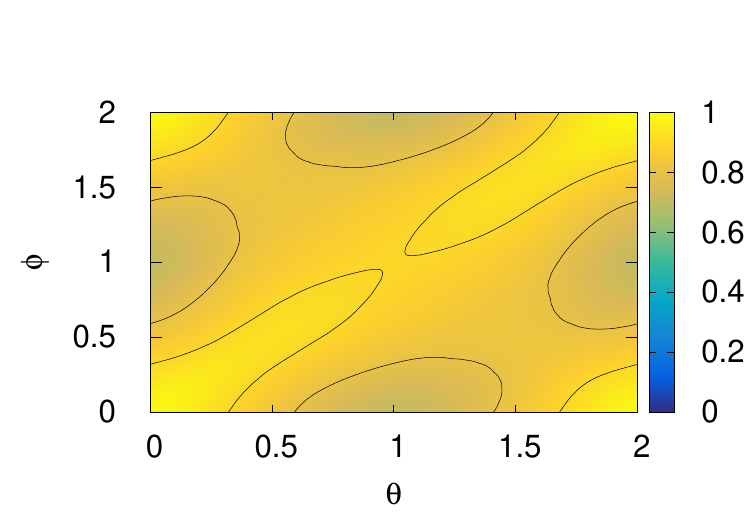}}\\
\subfigure[\label{fig:1d} ]{\includegraphics[scale=0.40]{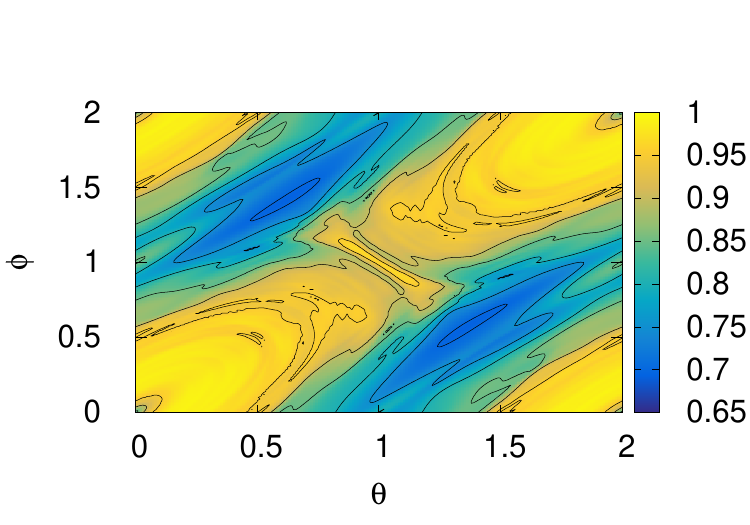}}\hspace*{.185cm}
\subfigure[\label{fig:1e} ]{\includegraphics[scale=0.40]{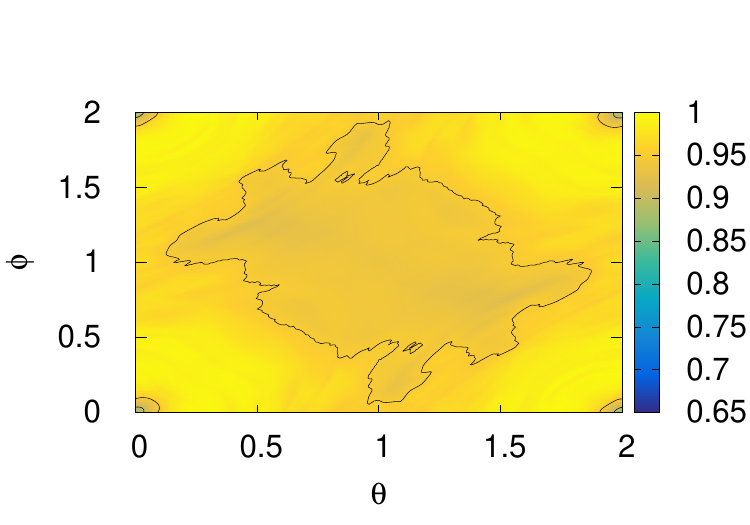}}\hspace*{.185cm}
\subfigure[\label{fig:1f} ]{\includegraphics[scale=0.40]{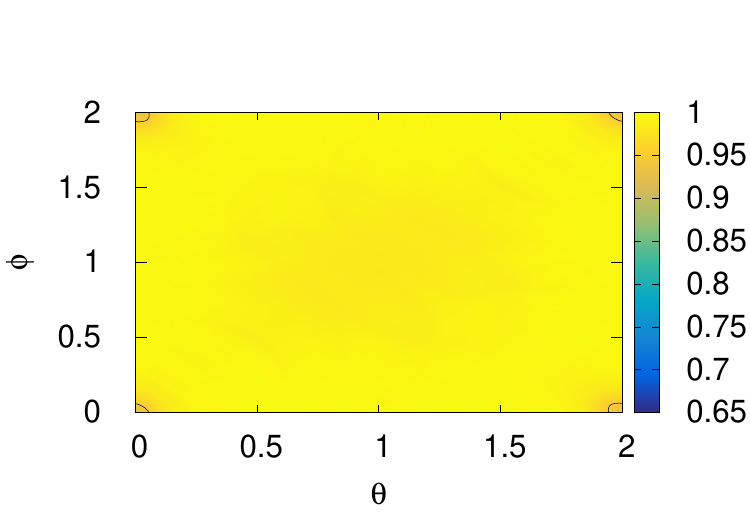}}\hspace*{0.00035cm}


\caption{\label{fig:1}{ Here figures (a)-(c) show the variations of the exponent \(\alpha(\theta,\phi)\) for the dynamic Fibonacci QW as a function of \(\theta,\phi\) respectively for \(\rho=0.2,0.5\mbox{ and }0.8\). Similarly, the figures (d)-(f) show the variations of the average entropy \(\langle S_{E}\rangle(\theta,\phi)  \) as a function of \(\theta,\phi\) respectively for \(\rho=0.2,0.5\mbox{ and }0.8\). Different colors have been used to indicate different values of \(\alpha\) and \(S_{E}\) as shown in the supplied color bars.  
 }
 }
\end{figure*}

\begin{figure}[th]
\centering

\subfigure[\label{fig:2a} ]{\includegraphics[scale=0.35]{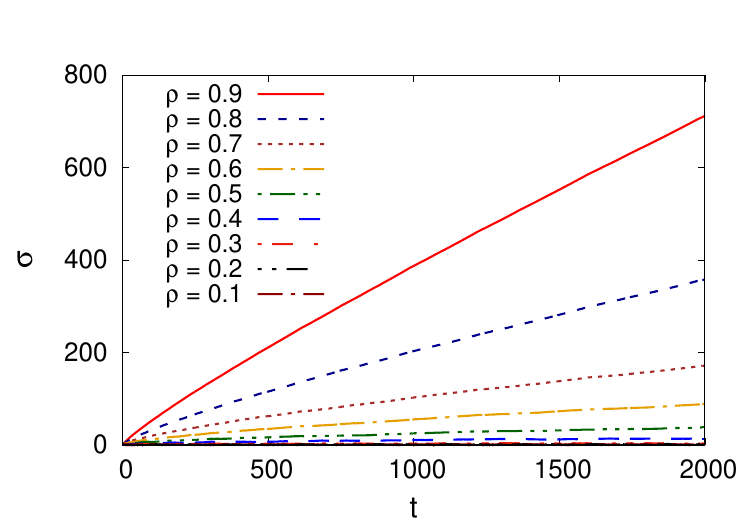}}\hspace*{-.035cm}
\subfigure[\label{fig:2b} ]{\includegraphics[scale=0.35]{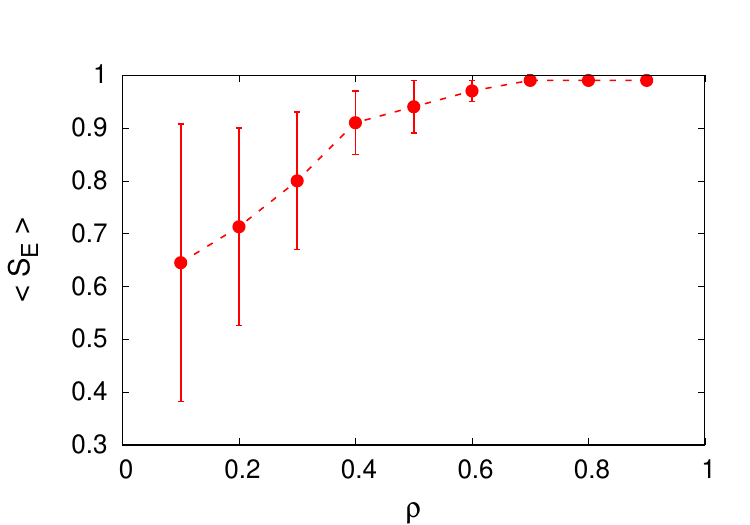}}

\caption{\label{fig:2}{ Here figure (a) shows the variations of standard deviation \(\sigma\) against time for different values of \(\rho\) in case of dynamic Fibonacci walk with the following coin parameters : \(\theta=0.67,\phi=1.49\). The figure (b) shows variation of \(\langle S_{E} \rangle \) against \(\rho\) for dynamic Fibonacci walk with the same set of parameters. 
 }
 }
\end{figure}

\section{ Results \& Discussions \label{three}}

We study the variations of the exponent \(\alpha(\rho_{1},\theta_{1},\phi_{1},\rho_{2},\theta_{2},\phi_{2})\) and the average entropy \(\langle S_{E}\rangle (\rho_{1},\theta_{1},\phi_{1},\rho_{2},\theta_{2},\phi_{2})\) as functions of the coin parameters for each deterministically aperiodic binary quantum walk. 
We study three cases for each walk : (i) \(\rho_{1}=\rho=0.2\), (ii) \(\rho_{1}=\rho=0.5\) and (iii) \(\rho_{1}=\rho=0.8\). For simplicity, we ignore the subscript of the parameters of \(C_{2}\). In each of the above three cases, we fix the \(\theta_{1},\phi_{1}\) parameters of \(C_{1}\) to \(\theta_{1}=0,\phi_{1}=0\) and then change the \(\theta,\phi\) parameters of \(C_{2}(\rho,\theta,\phi)\) in steps of \(\Delta \theta = 0.01\: \pi\) and \(\Delta \phi = 0.01\: \pi\) respectively. We demonstrate our results using contour plots of \(\alpha(\theta,\phi)\) and \(\langle S_{E}\rangle(\theta,\phi)\). The four points on the four corners of each such plot has identical values of the coin parameters i.e., \(\theta=\phi\) and therefore the values of \(\alpha\) and \(\langle S_{E}\rangle \) will be same as that for a conventional single coin walk.\\ 

In the following three subsections we discuss the results obtained for dynamic Fibonacci walk, the dynamic Rudin-Shapiro walk and the dynamic Thue-Morse walk. The results obtained for the static versions of the walks are described in the subsequent subsections. 
 \\

\subsection{Dynamic Fibonacci QW \label{d1}}
The dynamic Fibonacci walk exhibits the widest range of values of the exponent \(\alpha\). 
The contour plot in Fig. \ref{fig:1a} shows the variations of  \(\alpha\) as a function of (\(\theta,\phi\)) for  \(\rho=0.2\). It can be seen that there are different colored thick and thin elongated regions parallel to the diagonal \(\theta=\phi\). Different colors are used to indicate different values of \(\alpha\) as shown in the color bar provided in Fig. \ref{fig:1a}. 
The bright yellow colored elongated regions along the diagonal indicate strong sub-ballistic spreading of the walker. We find that nearly 29\% points exhibit strong sub-ballistic spreading with \(\alpha\) values in the range \(0.75 < \alpha < 1\). On the other hand, nearly same number of points exhibit localized behavior with corresponding \(\alpha\) values in the range \(-0.04 \leq \alpha < 0.25\) where -0.04 is the lowest value of \(\alpha\) exhibited by the system. The dark blue colored elongated regions on both sides of the diagonal indicate localization of the walker. The other thinner elongated regions, situated on both sides of the diagonal, of intermediate colors, indicates weak sub-ballistic, diffusive, sub-diffusive spreadings. The system also exhibits an wide range of values of the average entanglement entropy \(\langle S_{E} \rangle \). The variations of \(\langle S_{E} \rangle \) as a function of coin parameters is shown in \ref{fig:1d}. The bright yellow colored regions in fig. \ref{fig:1d}, near the two ends of the diagonal, indicate high values of \(\langle S_{E}\rangle \). On the contrary, the dark blue colored elongated regions on both sides of the diagonal indicate weak values of entropy. It is quite interesting to find that the region exhibiting low(high) spreading also exhibit smaller(higher) value of entropy. The minimum and maximum values of \(\langle S_{E} \rangle \) is \(0.69\pm 0.19\) and \(0.99\pm 0.01\). 
It can be seen from Fig. \ref{fig:1} that the values of \(\alpha\) and \(\langle S_{E}\rangle\) increase with increasing values of \(\rho\) at different \(\theta,\phi\) points with different rates. The contour plots in fig. \ref{fig:1b} and \ref{fig:1e} show the variations of \(\alpha\) and \(S_{E}\) respectively for \(\rho =0.5\). Comparing figs. \ref{fig:1a} and \ref{fig:1b}, we see that the dark blue region of fig. \ref{fig:1a} has turned more greenish in color and other such points have obtained more yellowish color. Only four small regions around the following four points  \((\pi,0), (0,\pi), (2\pi,\pi), (\pi,2\pi)\) exhibit subdiffusive spreading. The corresponding plot of entropy is shown in fig. \ref{fig:1e}. We find that nearly all points exhibits \(\langle S_{E} \rangle \) values in the \(0.90\leq \langle S_{E} \rangle < 1\). There is a significantly large patch of dark yellow color at the center. Many points in that region there exhibit \(\langle S_{E} \rangle \) in the range \(0.90 \leq \langle S_{E} \rangle \leq 0.95 \). The corner regions of the plot are colored in bright yellow indicating \(\langle S_{E} \rangle \) values in the range \(0.95<\langle S_{E} \rangle <1\). The plots in fig. \ref{fig:1c} and \ref{fig:1f} show the variations of \(\alpha\) and \(\langle S_{E}\rangle \) at  \(\rho = 0.8\). The color of the regions with strong sub-ballistic spreading(high average entanglement) in Fig.\ref{fig:1a}(\ref{fig:1c}) remain unchanged in Fig.\ref{fig:1b}(\ref{fig:1d}). Almost all points exhibit a sub-ballistic spreading along-with high degree of average entropy(\(\langle S_{E} \rangle > 0.95\)). Nearly 94 \% points exhibit \(\alpha\) values in the range \(0.75\leq \alpha < 1\). 1.17\%, 10\% and 73\% points exhibit \(\langle S_{E} \rangle \) values in the range \(0.99\leq \langle S_{E} \rangle < 1\) for \(\rho=0.2,0.5\mbox{ and } 0.8\) respectively.\\

The localization behavior found here in case of dynamic Fibonacci walks is quite interesting since, to the best of our knowledge, this is the first time that a localized behavior is found in case of any QW with dynamic binary disorder/aperiodicity. So far, localization has been reported only in case of binary static random disorder/aperiodicity.\\ 

In figure Fig. \ref{fig:2a}, we have shown the variation of \(\sigma\) against time for different values of \(\rho\) in case dynamic Fibonacci walk. It can be seen that the system exhibits localized behavior for \(\rho=0.1,0.2 \mbox{ and } 0.3\). At any time step, we find that \(\sigma\) gradually increases with increasing values of \(\rho\). It can be seen that if we increase \(\rho\), sub-diffusion, diffusion, sub-ballistic and nearly ballistic spreading can be obtained. On the other hand, in fig. Fig. \ref{fig:2b}, we have shown the variation of \(\langle S_{E} \rangle \) with \(\rho\). It can be seen that \(\langle S_{E} \rangle \) increases with increasing values of \(\rho\). For \(\rho \geq 0.7\), \(\langle S_{E} \rangle \) saturates to its maximal value. For lower values of \(\rho\), large error bars indicates that \(S_{E} \) fluctuates with time. The fluctuations reduce with increasing values of \(\rho\). These results indicate that the coin parameter \(\rho\) can be used as a tuning parameter to tune both the exponent \(\alpha\) and average entropy \(\langle S_{E} \rangle \) within significantly wide ranges for certain \(\theta,\phi\) values. Even a localization-delocalization transition can be realized by varying \(\rho\) at those points.  \\


It is also quite interesting to find that the region with low(high) spreading exhibit smaller(higher) value of entropy. In case of standard QW, ballistic spreading is found with weak entropy. On the other hand, fully disordered QW shows diffusive spreading with maximal entanglement \cite{q32}. The simple message was that dynamic disorder(of any degree) always increases entropy and decreases \(\alpha\). Here we find new information that nearly ballistic spreading can be obtained with maximal entanglement for the binary aperiodic QWs. The combination of aperiodicity and coin is playing the major role here. We can generate low \(\langle S_{E} \rangle \) with low \(\alpha\) or high \(\langle S_{E} \rangle\) with high \(\alpha\) using suitable combination of binary sequence and coin parameters. 

So, it is clear that the Fibonacci QW can exhibit a very versatile behavior. It is even more versatile than the behavior of QW with random disorder as reported in Ref.\cite{jing}. On the other hand, the ordered QW exhibits ballistic spreading independent of the parameters. So, the Fibonacci QW is much more sensitive to the coin parameters which is interesting as intuitively one would expect intermediate sensitiveness. \\
\begin{figure*}[h]
\centering

\subfigure[\label{fig:3a} ]{\includegraphics[scale=0.40]{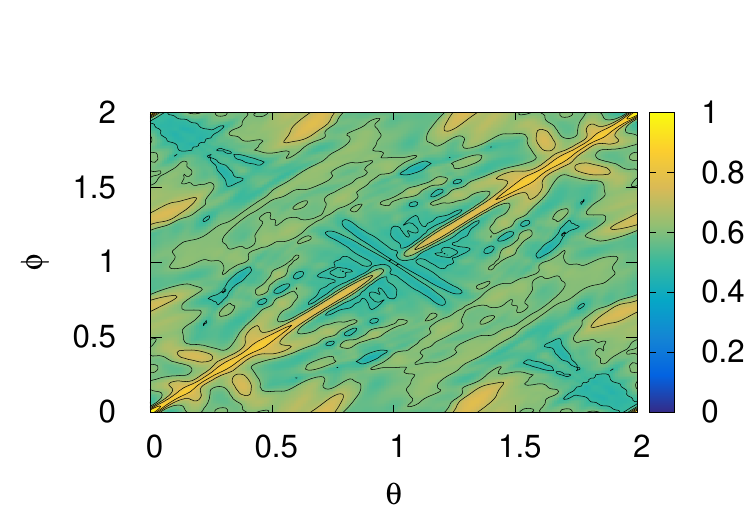}}\hspace*{.185cm}
\subfigure[\label{fig:3b} ]{\includegraphics[scale=0.40]{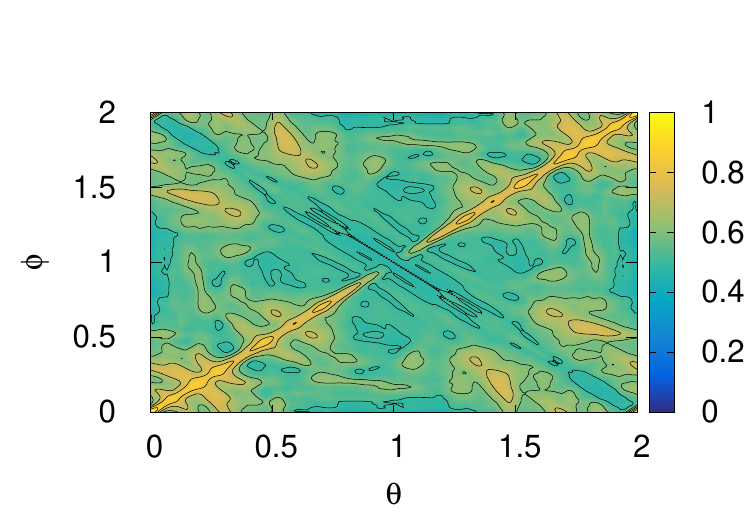}}\hspace*{.185cm}
\subfigure[\label{fig:3c} ]{\includegraphics[scale=0.40]{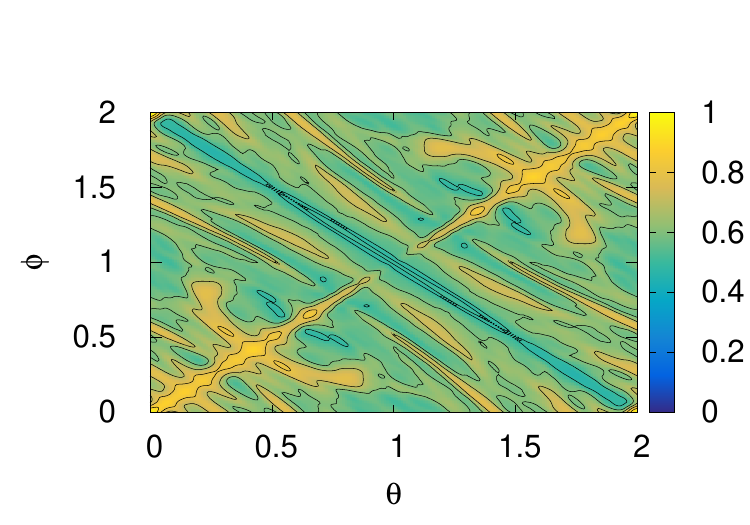}}\\
\subfigure[\label{fig:3d} ]{\includegraphics[scale=0.40]{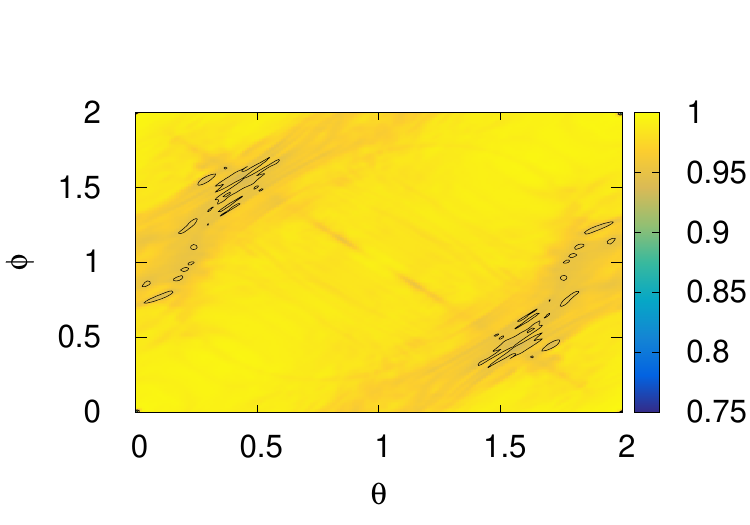}}\hspace*{.185cm}
\subfigure[\label{fig:3e} ]{\includegraphics[scale=0.40]{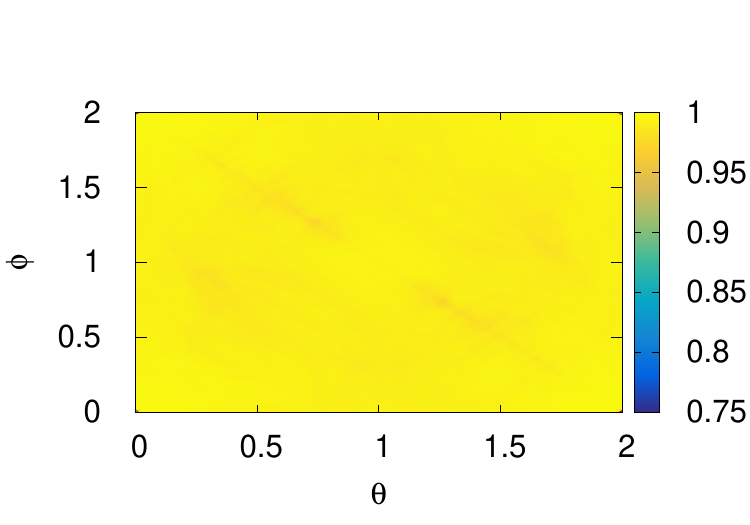}}\hspace*{.185cm}
\subfigure[\label{fig:3f} ]{\includegraphics[scale=0.40]{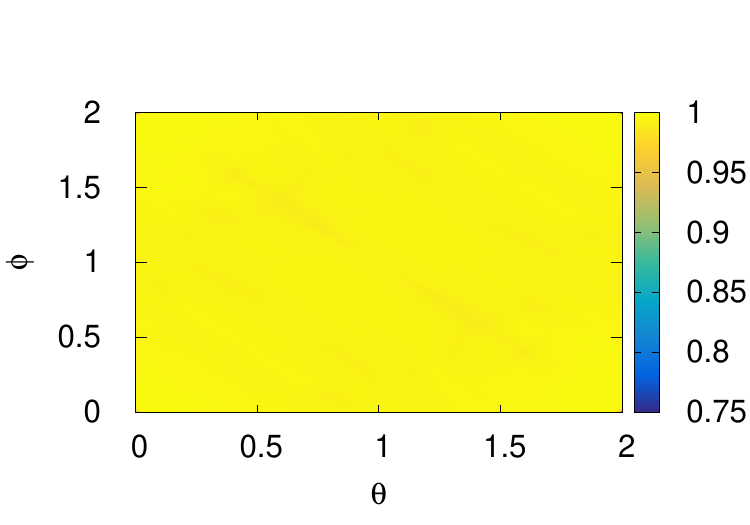}}\hspace*{.000185cm}


\caption{\label{fig:12}{ Here figures (a)-(c) show the variations of the exponent \(\alpha(\theta,\phi)\) for the dynamic Rudin-Shapiro QW as a function of \(\theta,\phi\) respectively for \(\rho=0.2,0.5\mbox{ and }0.8\). Similarly, the figures (d)-(f) show the variations of the average entropy \(\langle S_{E} \rangle (\theta,\phi)\) as a function of \(\theta,\phi\) respectively for \(\rho=0.2,0.5\mbox{ and }0.8\). Different colors have been used to indicate different values of \(\alpha\) and \(S_{E}\) as shown in the supplied color bars.  
 }
 }
\end{figure*}
\begin{figure*}[h]
\centering

\subfigure[\label{fig:4a} ]{\includegraphics[scale=0.40]{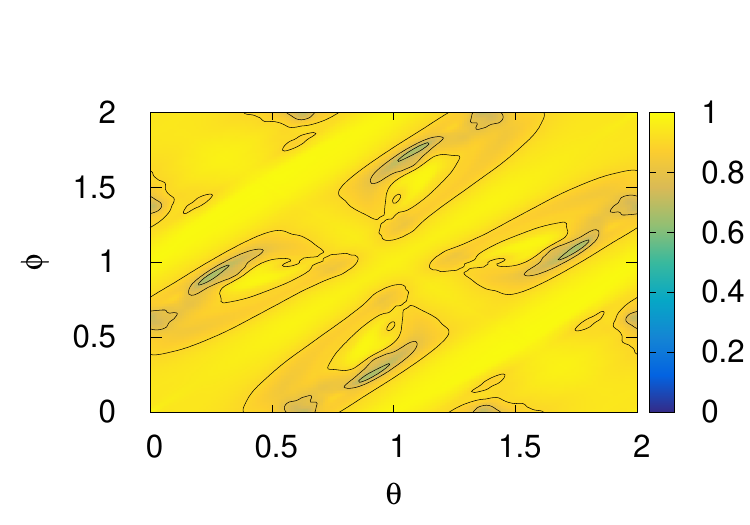}}\hspace*{.185cm}
\subfigure[\label{fig:4b} ]{\includegraphics[scale=0.40]{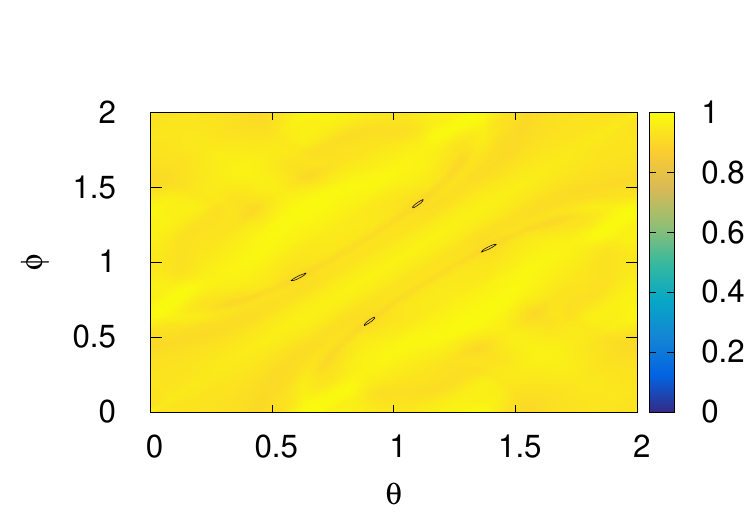}}\hspace*{.185cm}
\subfigure[\label{fig:4c} ]{\includegraphics[scale=0.40]{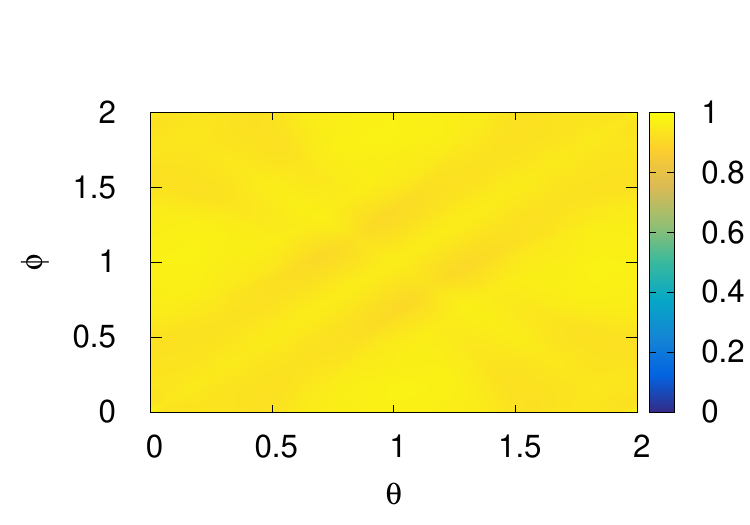}}\\
\subfigure[\label{fig:4d} ]{\includegraphics[scale=0.40]{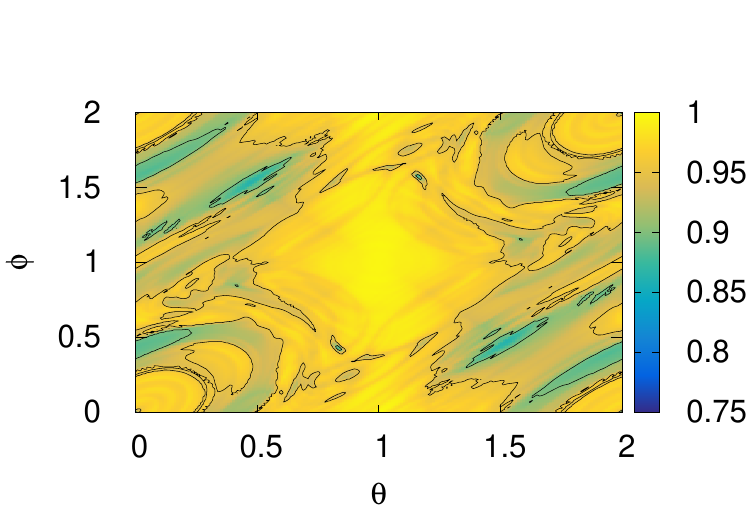}}\hspace*{.185cm}
\subfigure[\label{fig:4e} ]{\includegraphics[scale=0.40]{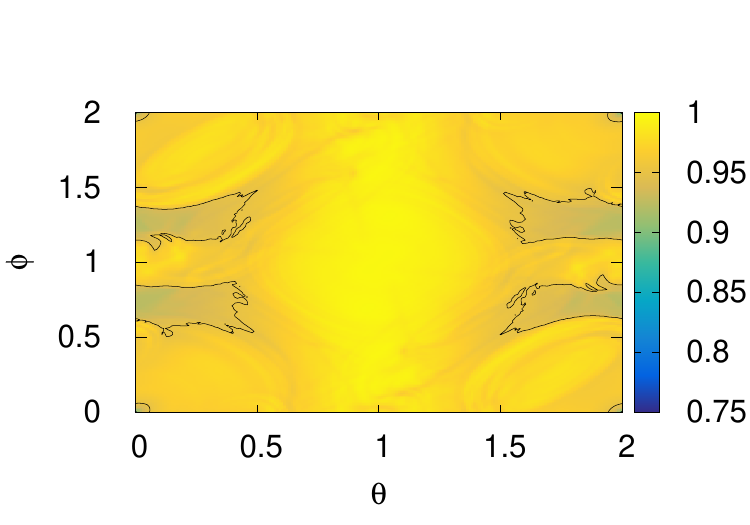}}\hspace*{.185cm}
\subfigure[\label{fig:4f} ]{\includegraphics[scale=0.40]{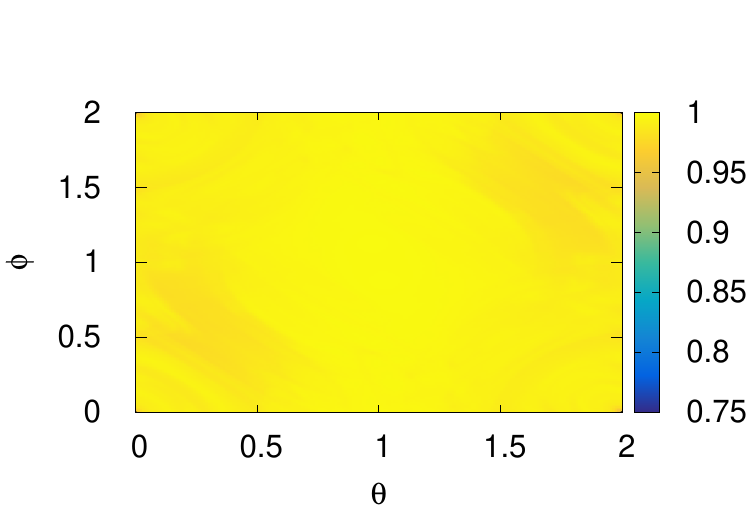}}\hspace*{.000185cm}


\caption{\label{fig:13}{ Here figures (a)-(c) show the variations of the exponent \(\alpha(\theta,\phi)\) for the dynamic Thue-Morse QW as a function of \(\theta,\phi\) respectively for \(\rho=0.2,0.5\mbox{ and }0.8\). Similarly, the figures (d)-(f) show the variations of the average entropy \(\langle S_{E} \rangle (\theta,\phi)\) as a function of \(\theta,\phi\) respectively for \(\rho=0.2,0.5\mbox{ and }0.8\). Different colors have been used to indicate different values of \(\alpha\) and \(S_{E}\) as shown in the supplied color bars.  
 }
 }
\end{figure*}

\subsection{Dynamic Rudin-Shapiro QW: \label{d2}}

The dynamic Rudin-Shapiro walk exhibits a spreading behavior which is much less sensitive to coin parameter variation in comparison to the dynamic Fibonacci walk. The system mainly exhibits sub-diffusive, diffusive and weak sub-ballistic spreadings along with high degree of entropy for \(\rho=0.2,0.5 \mbox{ and }0.8\). Figs. \ref{fig:3a}, \ref{fig:3b} and \ref{fig:3c} show the variations of \(\alpha\) for \(\rho=0.2,0.5 \mbox{ and }0.8\) respectively. All the three contour plots largely show colors indicating sub-diffusion,diffusion and weak sub-ballistic behavior. We find that 92.4\%, 92.1\%, 82.2\% points lie between 0.40 to 0.70 respectively for \(\rho =0.2,0.5 \mbox{ and }0.8\) i.e., majority of (\(\theta,\phi\)) points exhibit \(\alpha\) values in this range and it is somewhat independent of \(\rho\) . However, shapes of different colored regions  change with \(\rho\) as shown in the figure. The light greenish color background indicate diffusion in all the three figures. It can be seen from all the three figures that there are multiple bluish green colored regions of various shapes and sizes. The walker exhibit sub-diffusive spreading at those regions.  Similarly dark yellow colored regions of different shapes and sizes are present in all three plots. The walker exhibits sub-ballistic spreading at those regions. One distinct feature is that sub-diffusive spreading is obtained throughout the other diagonal for \(\rho=0.8\). Only 0.07\%, 0.09\% and 0.11\% points exhibit \(\alpha\) values between 0.9 and 1 respectively for \(\rho=0.2,0.5\mbox{ and }0.8\). These minority points are placed along the \(\theta=\phi\) diagonal. On the other hand, all the three contour plots in the figs. \ref{fig:4d}, \ref{fig:4e} and \ref{fig:4f} show the walker predominantly exhibit highly entangled behavior for different values of the coin parameters. Most part of Fig. \ref{fig:4d} is colored in bright yellow indicating highly entangled dynamics. We find that nearly 98\% points exhibit \(\langle S_{E} \rangle \) values greater than 0.95. The other points exhibit \(\langle S_{E} \rangle \) values lying in the range \(0.90\leq \langle S_{E} \rangle < 0.95 \). These points are positioned on the dark yellow colored regions of the plot. For higher values of \(\rho\), more than 99\% points exhibit \(\langle S_{E} \rangle \) values in the range \(0.95\leq \langle S_{E} \rangle < 1 \). Around 19.5\%, 74\% and 99\% points exhibit \( \langle S_{E} \rangle  \) values \(>0.99\) respectively for \(\rho=0.2,0.5 \mbox{ and }0.8\). All these points are strong candidates for exhibiting maximum entanglement.\\ As shown in the sections \ref{d1} \& \ref{d3}, the Fibonacci and the Thue-Morse walks exhibit predominantly sub-ballistic spreading at \(\rho=0.8\). This  difference separates binary Rudin-Shapiro walk from the other two walks.\\ The spreading behavior of dynamic Rudin-Shapiro sequence in response to coin parameter variation is quite different from that of random disorder case which did not exhibit sub-diffusive spreading according to ref.\cite{jing}.\\

\begin{figure*}[h]
\centering
\subfigure[\label{fig:5a} ]{\includegraphics[scale=0.40]{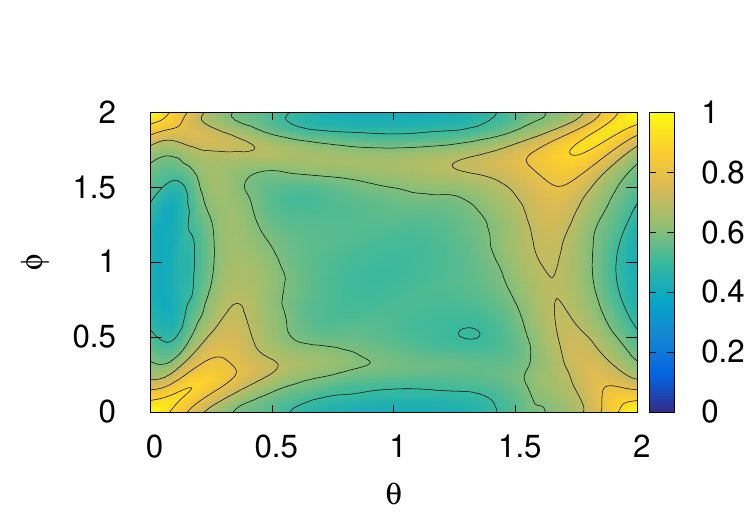}}\hspace*{.185cm}
\subfigure[\label{fig:5b} ]{\includegraphics[scale=0.40]{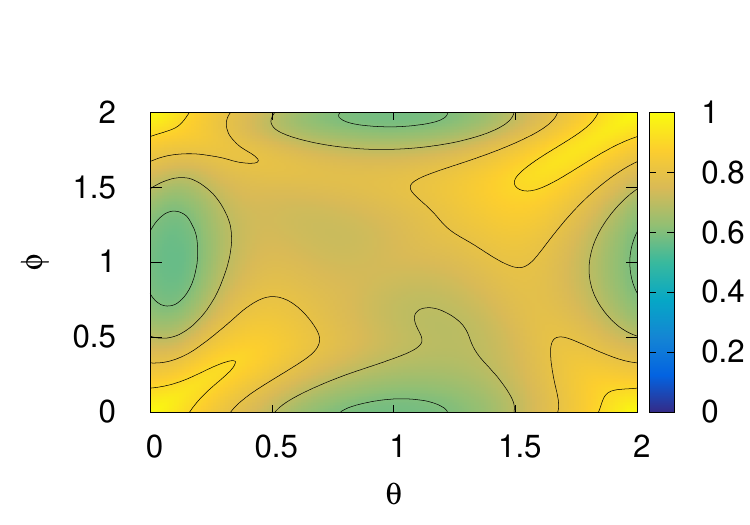}}\hspace*{.185cm}
\subfigure[\label{fig:5c} ]{\includegraphics[scale=0.40]{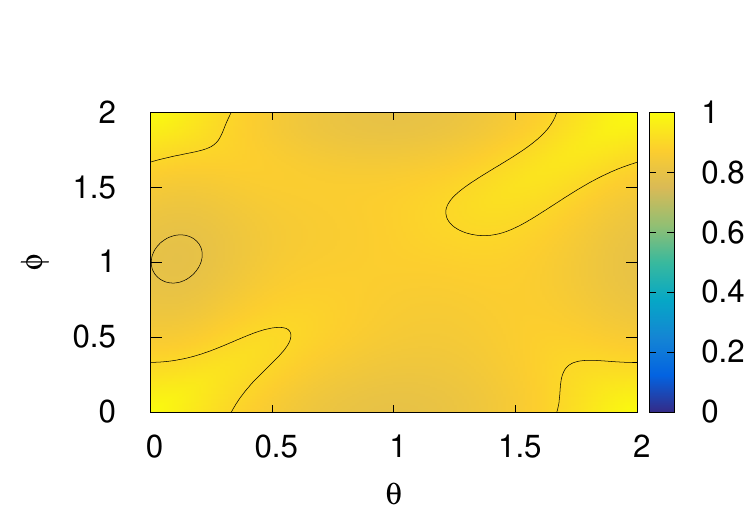}}\\
\subfigure[\label{fig:5d} ]{\includegraphics[scale=0.40]{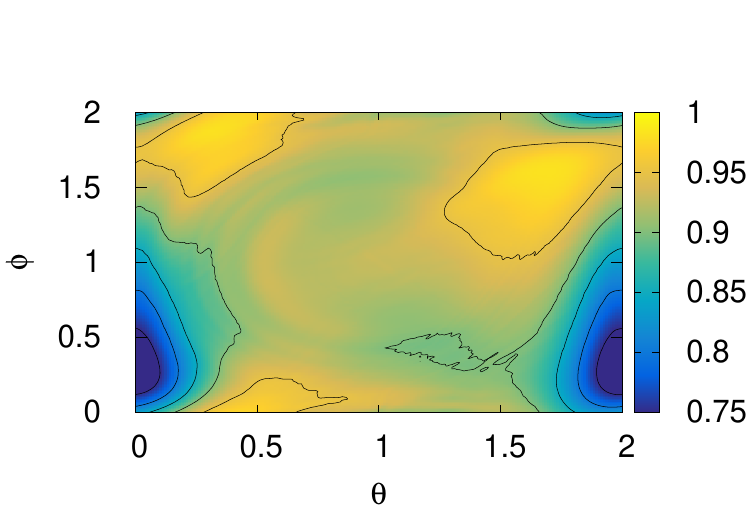}}\hspace*{.185cm}
\subfigure[\label{fig:5e} ]{\includegraphics[scale=0.40]{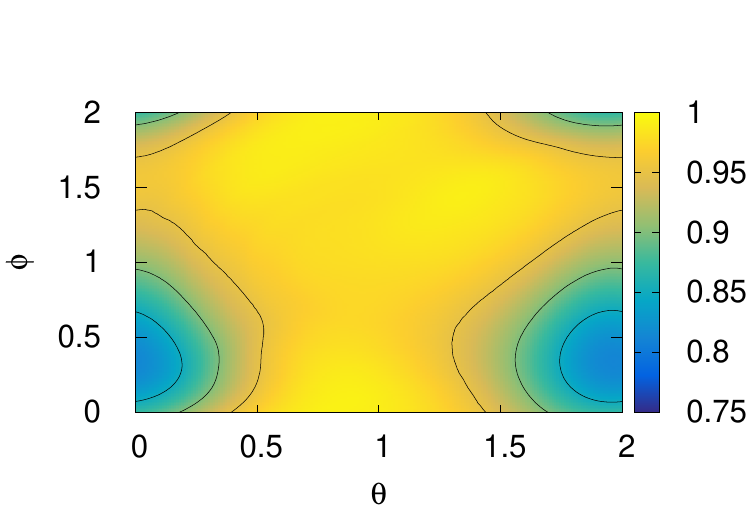}}\hspace*{.185cm}
\subfigure[\label{fig:5f} ]{\includegraphics[scale=0.40]{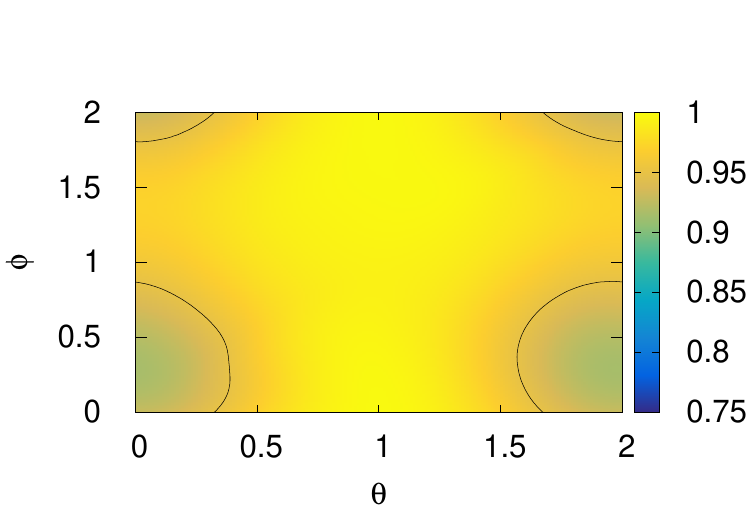}}\hspace*{.000185cm}

\caption{\label{fig:11}{ Here figures (a)-(c) show the variations of the exponent \(\alpha(\theta,\phi)\) for the static Fibonacci QW as a function of \(\theta,\phi\) respectively for \(\rho=0.2,0.5\mbox{ and }0.8\). Similarly, the figures (d)-(f) show the variations of the average entropy \(\langle S_{E}\rangle(\theta,\phi)\) as a function of \(\theta,\phi\) respectively for \(\rho=0.2,0.5\mbox{ and }0.8\). Different colors have been used to indicate different values of \(\alpha\) and \(S_{E}\) as shown in the supplied color bars.  
 }
 }
\end{figure*}

\subsection{Dynamic Thue-Morse QW \label{d3}:} 
Dynamic Thue-Morse QW exhibits a spreading behavior which is the least sensitive to the coin parameter variation among the three different dynamic QWs studied here. It predominantly exhibits strong sub-ballistic behavior alongwith high values of entanglement. 
The contour plots in figs. \ref{fig:4a}, \ref{fig:4b} and \ref{fig:4c} show the variations of \(\alpha\) for \(\rho=0.2,0.5 \mbox{ and }0.8\) respectively in case of dynamic Thue-Morse walk. It can be seen that most part of the plot in Fig. \ref{fig:4a} is colored in bright yellow which indicate strong sub-ballistic spreading. We have found that nearly 98\% of the points exhibit \(\alpha\) values in the range \(0.75 < \alpha < 1\). The remaining points, exhibit weak sub-ballistic spreading (\(0.5 < \alpha \leq 0.75\)). It is worth mentioning that none of the other aperiodic QWs studied here exhibit such predominant strong sub-ballistic behavior for \(\rho=0.2\). It also offers the shortest range of \(\alpha\) values for \(\rho=0.2\) in comparison to the others. For higher values of \(\rho\), strong sub-ballistic spreading is found at all points as indicated in figs. \ref{fig:4b} and \ref{fig:4c}. Most part of these plots are colored in bright yellow. We find that almost all points exhibit \(\alpha\) values lying between between 0.90 and 1 for both \(\rho=0.5\mbox{ and }0.8\). For \(\rho=0.5\), the lowest value of \(\alpha\) is 0.90. We also find that 3.53, 2.0 and 0.01 are the percentages of points having \(\alpha\geq 0.99\) respectively for \(\rho=0.2,0.5 \mbox{ and }0.8\). So, a strange thing to note here is that at some points \(\alpha\) value decreases with increasing values of \(\rho\). Such behavior was not found for other walks studied here. The system exhibits \(\langle S_{E} \rangle \) values in the ranges \(0.83\leq \langle S_{E}\rangle < 1\), \(0.97\leq \langle S_{E}\rangle < 1\) and \(0.93\leq \langle S_{E}\rangle < 1\) respectively for \(\rho=0.2,0.5 \mbox{ and }0.8\). Figs. \ref{fig:4d}, \ref{fig:4e} and \ref{fig:4f} show the corresponding variations of \(\langle S_{E}\rangle \) as a function of \(\theta,\phi\) respectively for \(\rho=0.2,0.5 \mbox{ and }0.8\). It can be seen from the figures that \(\langle S_{E} \rangle \) is most sensitive to \(\theta,\phi\) variation in case of Fig. \ref{fig:4d}. The plot in Fig. \ref{fig:4d} shows that there are multiple bluish green colored regions of various shapes and sizes. We find that those regions mainly consists of around 4\% points which exhibit \(\langle S_{E}\rangle\) values below 0.90. The dark yellow colored regions cover more parts of the plot. These regions mainly consists of points exhibiting \(\langle S_{E}\rangle\) values lying between 0.90 and 0.95. The bright yellow colored regions exhibit \(\langle S_{E}\rangle\) values lying between 0.95 and 1. For \(\rho=0.5\), most part of the plot is colored in bright yellow. We find that around 84\% points exhibit \(\langle S_{E}\rangle\) values greater than 0.95. Around 15\% points exhibit \(\langle S_{E}\rangle\) values lying between 0.90 and 0.95, most of which fall in the four distinct dark yellow colored regions. For \(\rho=0.8\), \(\langle S_{E}\rangle > 0.95\) at all points. 3.7\%, 13.8\% and 73\% points exhibit \(\langle S_{E} \rangle \) values in the range \(0.99\leq \langle S_{E} \rangle < 1\) for \(\rho=0.2,0.5\mbox{ and } 0.8\) respectively.\\


\begin{figure*}[h]
\centering
\subfigure[\label{fig:6a} ]{\includegraphics[scale=0.40]{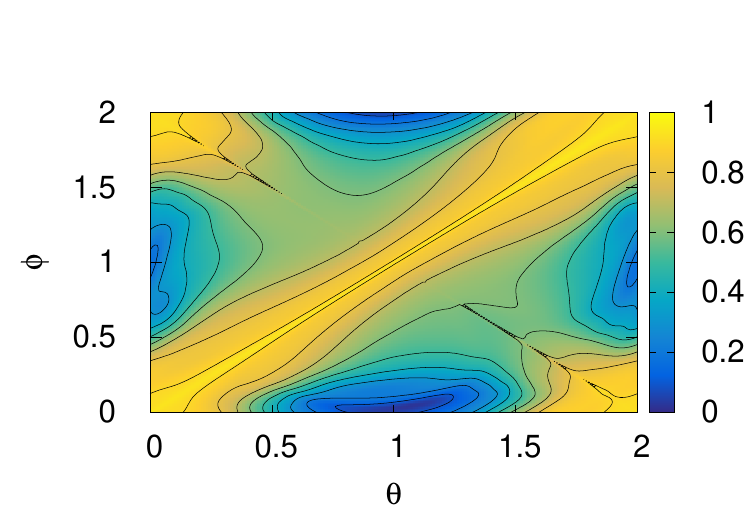}}\hspace*{.185cm}
\subfigure[\label{fig:6b} ]{\includegraphics[scale=0.40]{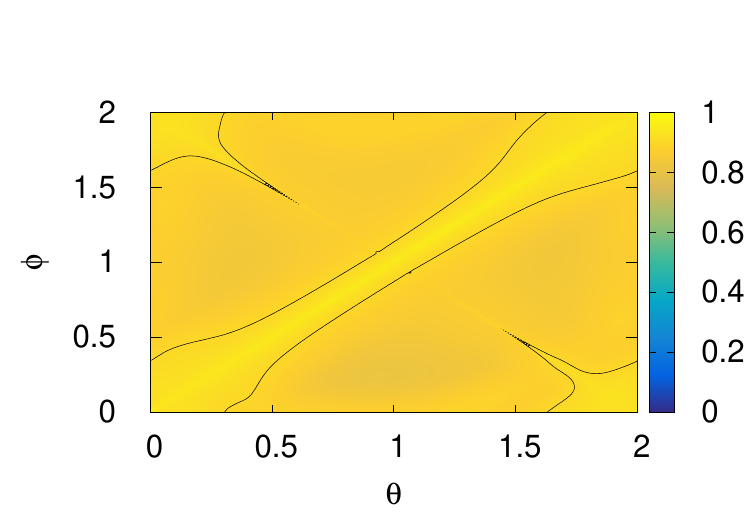}}\hspace*{.185cm}
\subfigure[\label{fig:6c} ]{\includegraphics[scale=0.40]{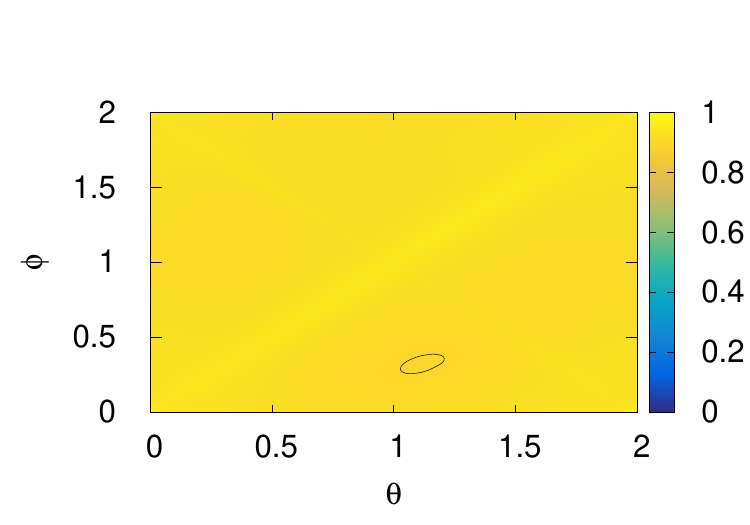}}\\
\subfigure[\label{fig:6d} ]{\includegraphics[scale=0.40]{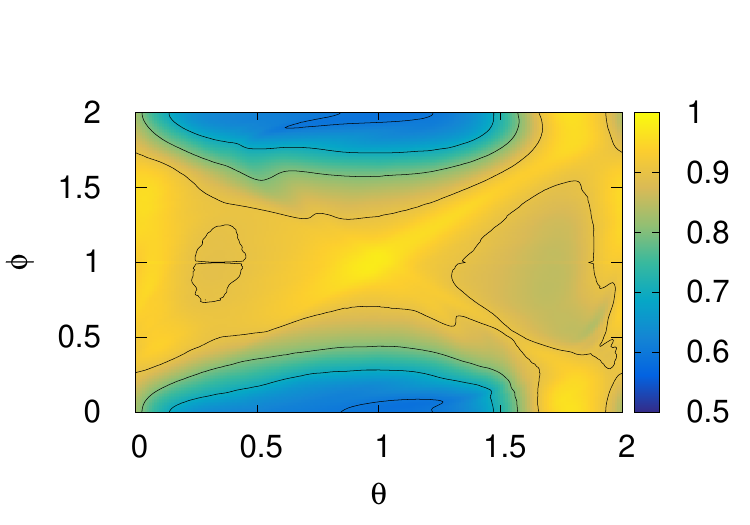}}\hspace*{.185cm}
\subfigure[\label{fig:6e} ]{\includegraphics[scale=0.40]{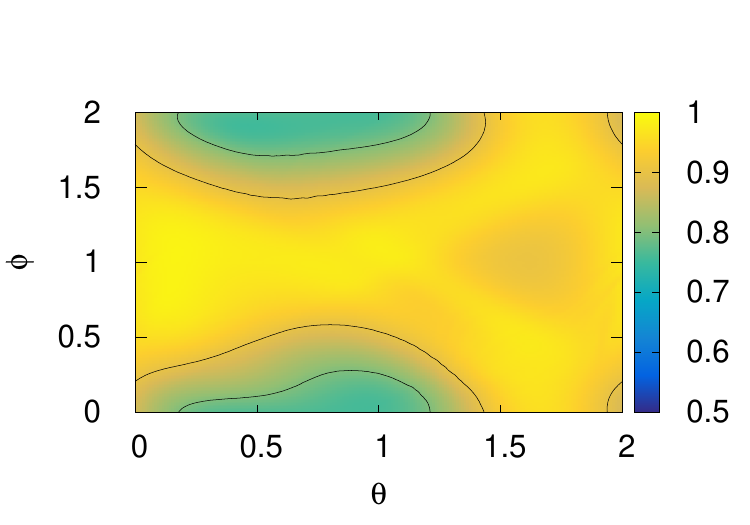}}\hspace*{.185cm}
\subfigure[\label{fig:6f} ]{\includegraphics[scale=0.40]{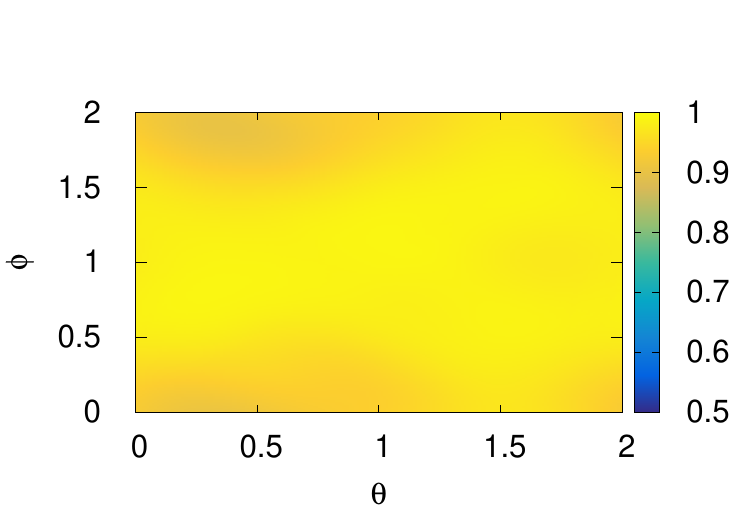}}\hspace*{.0000185cm}

\caption{\label{fig:13}{Here figures (a)-(c) show the variations of the exponent \(\alpha(\theta,\phi)\) for the static Thue-Morse QW as a function of \(\theta,\phi\) respectively for \(\rho=0.2,0.5\mbox{ and }0.8\). Similarly, the figures (d)-(f) show the variations of the average entropy \(\langle S_{E}\rangle (\theta,\phi)\) as a function of \(\theta,\phi\) respectively for \(\rho=0.2,0.5\mbox{ and }0.8\). Different colors have been used to indicate different values of \(\alpha\) and \(S_{E}\) as shown in the supplied color bars.  
 }
 }
\end{figure*}
\begin{figure*}[h]
\centering

\subfigure[\label{fig:7a} ]{\includegraphics[scale=0.40]{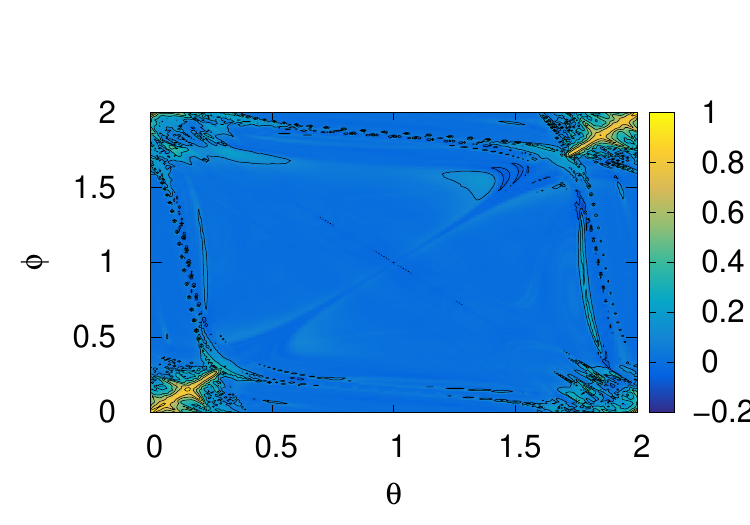}}\hspace*{.185cm}
\subfigure[\label{fig:7b} ]{\includegraphics[scale=0.40]{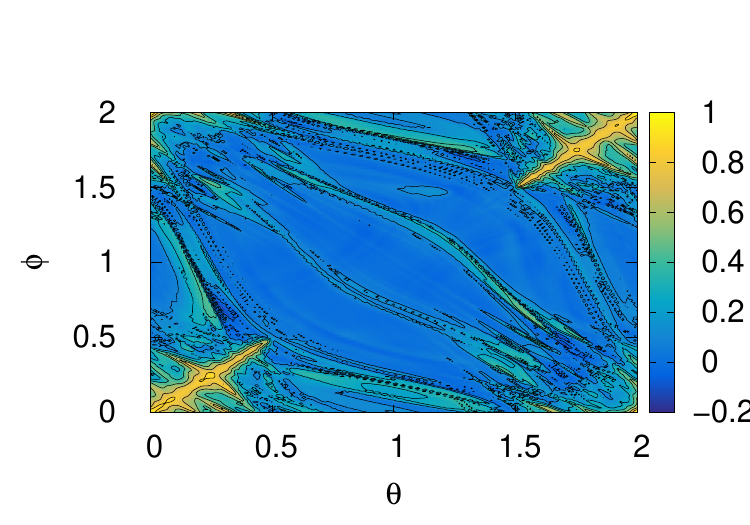}}\hspace*{.185cm}
\subfigure[\label{fig:7c} ]{\includegraphics[scale=0.40]{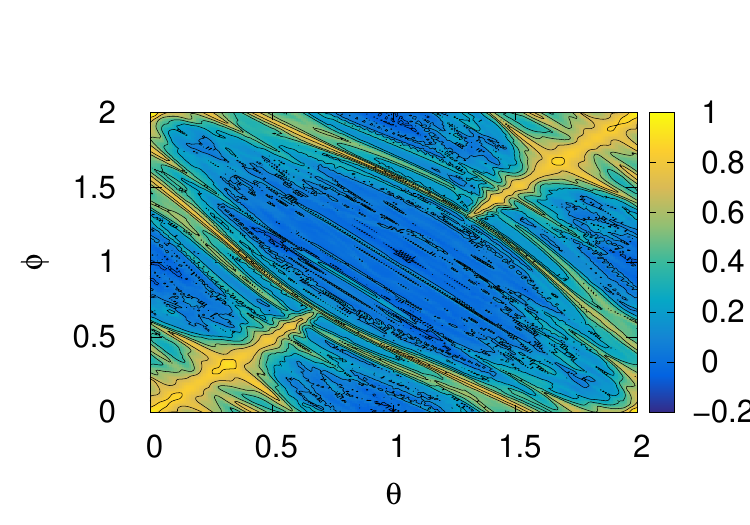}}\\
\subfigure[\label{fig:7d} ]{\includegraphics[scale=0.40]{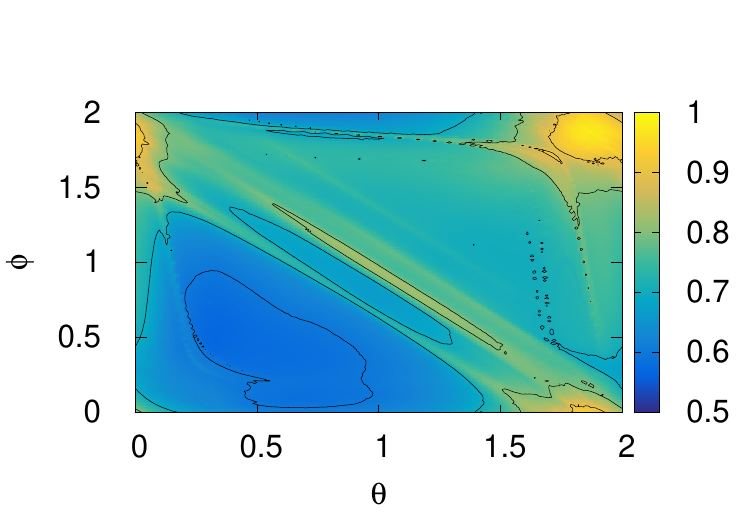}}\hspace*{.185cm}
\subfigure[\label{fig:7e} ]{\includegraphics[scale=0.40]{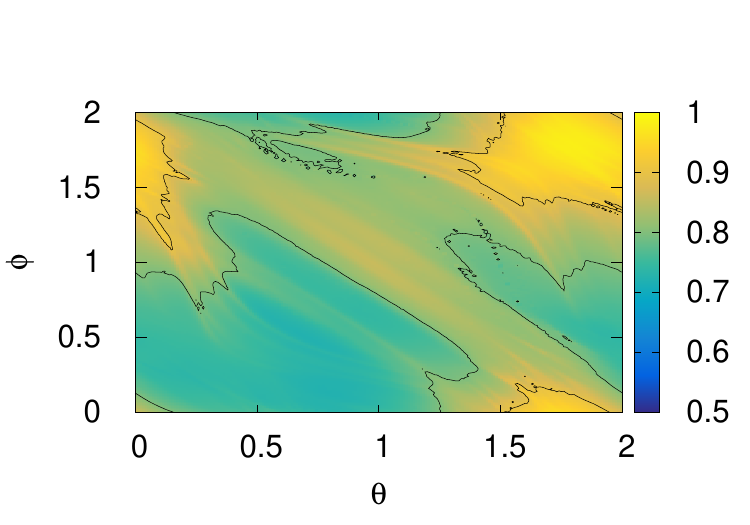}}\hspace*{.185cm}
\subfigure[\label{fig:7f} ]{\includegraphics[scale=0.40]{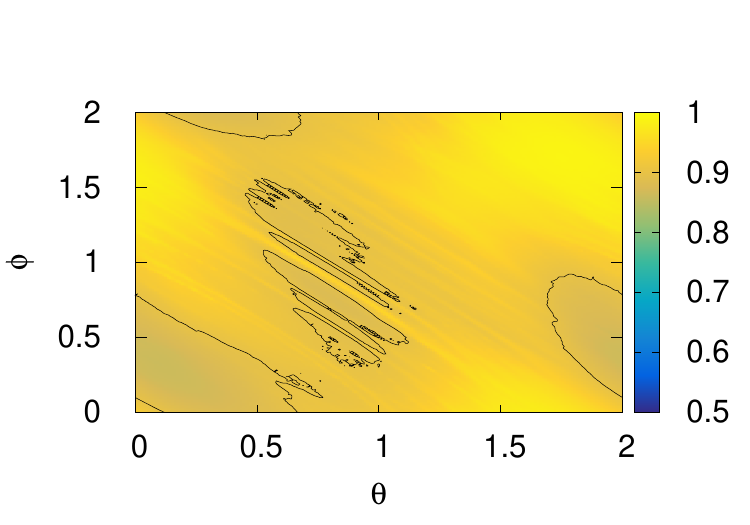}}\hspace*{.0005cm}

\caption{\label{fig:12}{ Here figures (a)-(c) show the variations of the exponent \(\alpha(\theta,\phi)\) for the static Rudin-Shapiro QW as a function of \(\theta,\phi\) respectively for \(\rho=0.2,0.5\mbox{ and }0.8\). Similarly, the figures (d)-(f) show the variations of the average entropy \(\langle S_{E}\rangle(\theta,\phi)\) as a function of \(\theta,\phi\) respectively for \(\rho=0.2,0.5\mbox{ and }0.8\). Different colors have been used to indicate different values of \(\alpha\) and \(S_{E}\) as shown in the supplied color bars.  
 }
 }
\end{figure*}

\begin{figure*}[h]
\centering

\subfigure[\label{fig:8a} ]{\includegraphics[scale=0.25]{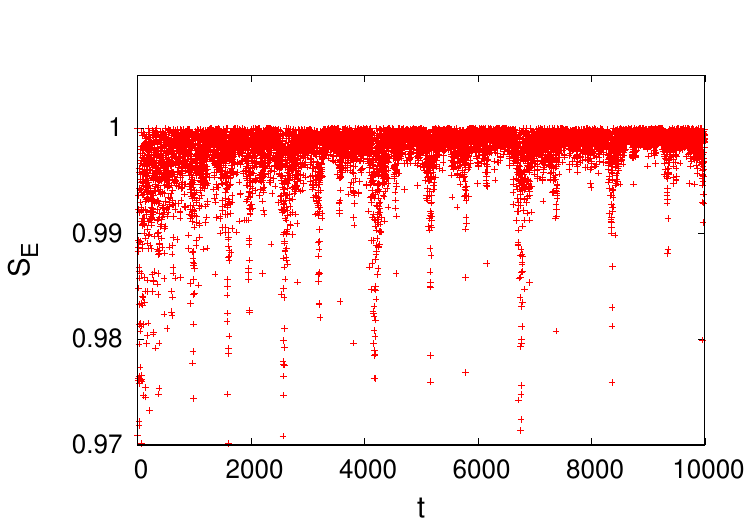}}\hspace*{-.035cm}
\subfigure[\label{fig:8b} ]{\includegraphics[scale=0.25]{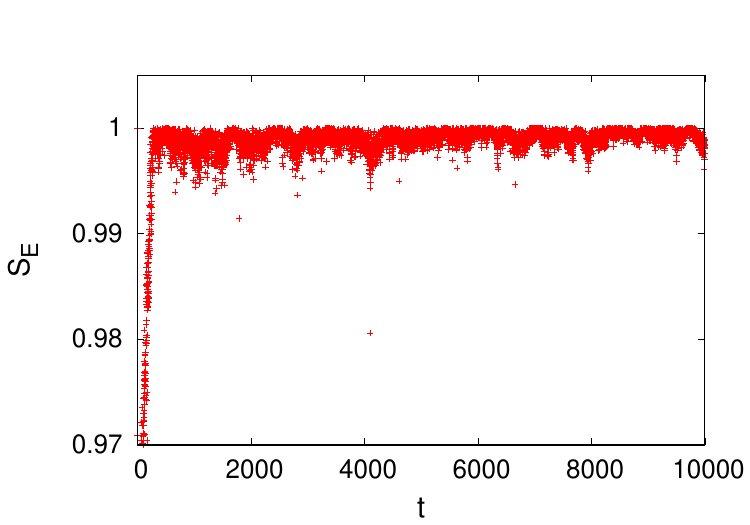}}\hspace*{-.035cm}
\subfigure[\label{fig:8c} ]{\includegraphics[scale=0.25]{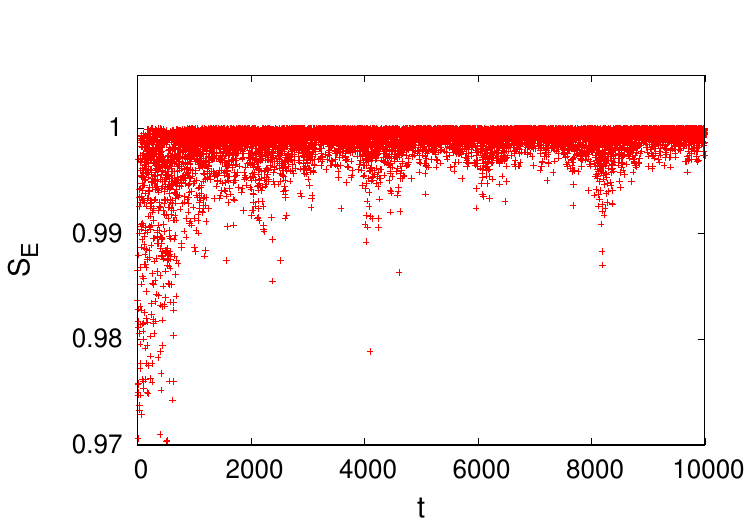}}\hspace*{-.035cm}
\subfigure[\label{fig:8d} ]{\includegraphics[scale=0.25]{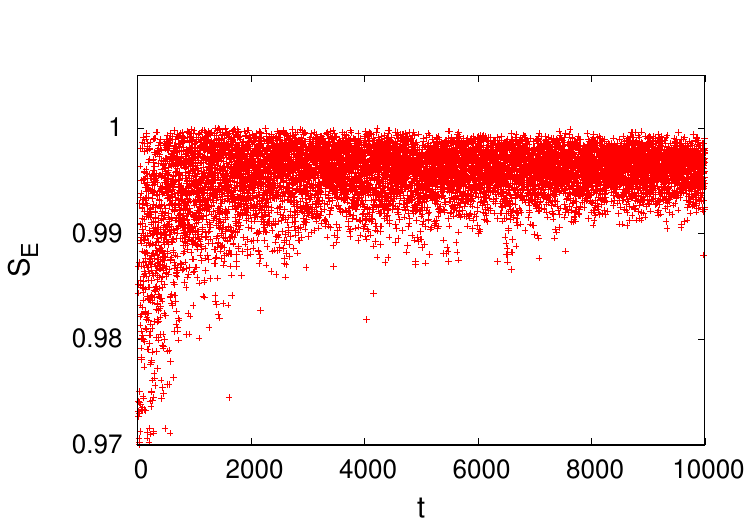}}\hspace*{-.035cm}
\subfigure[\label{fig:8e} ]{\includegraphics[scale=0.25]{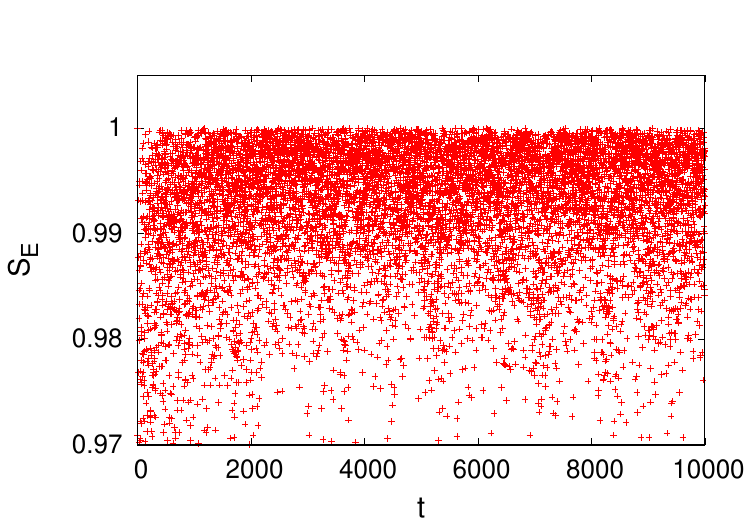}}

\caption{\label{fig:13}{ (a) Variation of entropy \(S_{E}\) against time for the dynamic Fibonacci walk at \(\rho = 0.8,\theta = 0.03  \pi , \phi = 1.46 \pi \), (b) Variation of entropy \(S_{E}\) against time for the dynamic Rudin-Shapiro walk at \(\rho = 0.8,\theta = 0 , \phi = 0.08 \pi \), (c) Variation of entropy \(S_{E}\) against time for the dynamic Thue Morse walk at \(\rho = 0.8,\theta = 1.06  \pi , \phi = 0.80 \pi \), (d) Variation of entropy \(S_{E}\) against time for the static Fibonacci walk at \(\rho = 0.8,\theta = 1.06  \pi , \phi = 0.80 \pi \), (e) Variation of entropy \(S_{E}\) against time for the static Thue-Morse walk at \(\rho = 0.8,\theta = 0.90  \pi , \phi = 1.41 \pi \),   
 }
 }
\end{figure*}

\subsection{Static Fibonacci Walk \label{s1}:} Figs. \ref{fig:5a}, \ref{fig:5b} and \ref{fig:5c} show the variations of \(\alpha\) for \(\rho=0.2,0.5\mbox{ and }0.8\) respectively in case of static Fibonacci walk. One can see from Fig. \ref{fig:5a} that most part of the plot has colors indicating sub-diffusion, diffusion and weak sub-ballistic spreadings. We find that around 90\% points exhibit \(\alpha\) values in the range \(0.40\leq \alpha \leq 0.75\). There are four distinct elongated bluish green colored regions on four different sides of the plot which exhibit sub-diffusive spreadings. Two of them are thin \& elongated in shape, oriented parallel to the \(\theta\) axis, and are situated around the points (\(\pi,0\)) and (\(\pi,2\pi\)). The other two regions are situated around the points (\(0,\pi\)) and (\(2\pi,\pi\)). The greenish color in the central region of the plot indicate weak sub-ballistic spreading is found there. There are some points near the four corners of the plot which exhibit strong sub-ballistic spreading (\(\alpha\geq 0.75\)). Figs. \ref{fig:5d}, \ref{fig:5e} and \ref{fig:5f} show the variations of \(\langle S_{E}\rangle \) for \(\rho=0.2,0.5\mbox{ and } 0.8\) respectively. The system exhibits \(\langle S_{E} \rangle \) values in the range \(0.71\leq \langle S_{E} \rangle < 0.98\) for \(\rho=0.2\). It can be seen from Fig. \ref{fig:5a} that there are few regions of yellow color near the top and bottom parts of the contour plot which exhibits higher values of \(\langle S_{E} \rangle \) whereas the middle part exhibits lower values of \(\langle S_{E} \rangle \). There is one bright yellow colored region near the bottom left side of the plot which exhibit high entropy. We find that only 16\% points exhibit \(\langle S_{E} \rangle \) values in the range \(S_{E}>0.95\). For higher values of \(\rho\), the system exhibits weak and strong sub-ballistic spreadings. The four regions which exhibited sub-diffusive spreadings for \(\rho=0.2\), exhibits weak sub-ballistic spreading for \(\rho=0.5\). Around 26\% points exhibit \(\alpha\) values in the range \(0.8\leq \alpha \leq 0.95\). For \(\rho=0.8\), light and dark yellow color throughout the plot indicates strong sub-ballistic spreading at all points. Around 18\% points exhibit \(\alpha\) values in the range (\(0.9\leq \alpha < 1\)). Entropy values at different points also increase with increasing values of \(\rho\). Bright yellow color can be seen in most parts of the figure. The range of \(\langle S_{E} \rangle \) values changes to \(0.81\leq \langle S_{E} \rangle \leq .993\) and \(0.92\leq \langle S_{E} \rangle \leq .998\) respectively for \(\rho=0.5\) and 0.8. Lower entropy is found in the regions around the four corners. 3.5\% and 30.6\% points exhibit \(\langle S_{E} \rangle \) values in the range \(0.99\leq \langle S_{E} \rangle < 1\) for \(\rho=0.5\mbox{ and } 0.8\) respectively. These points are possible candidates for exhibiting maximal entanglement. It is interesting to note that we found localization in dynamic Fibonacci walk but static Fibonacci walk does not exhibit a localization phenomenon. This indicates that, at some (\(\theta,\phi\)) points, static disorder generates faster spreading than dynamic disorder. One more point to note is the similarity with dynamic walks that \(\alpha\) and \(\langle S_{E}\rangle\) are more sensitive to (\(\theta,\phi\)) variation at \(\rho=0.2\) in comparison to the other values of \(\rho\).\\ 



\subsection{Static Thue-Morse Walk :\label{s2}} The variation of the spreading exponent are shown in Figs.\ref{fig:6a}, \ref{fig:6b} and \ref{fig:6c} respectively for \(\rho=0.2, 0.5\mbox{ and }0.8\). It can be seen that strong sub-ballistic spreading is there along the diagonal \(\theta=\phi\) for all three values of \(\rho\). Existence of various colors in the contour plot shown in Fig.\ref{fig:6a} indicate that the system exhibits various types of spreading from localized to strong sub-ballistic spreading for \(\rho=0.2\). Nearly 23\% points exhibit localized or sub-diffusive behavior. It can be seen from Fig. \ref{fig:6a} that such points mainly lie inside the four dark blue colored patches around the following four points (\(\pi,0\)), (\(\pi,2\pi\)), (\(0,\pi\)) and (\(2 \pi,\pi\)). Only 5\% points exhibit \(\alpha\) values below \( 0.25\) and the minimum found value of the exponent is 0.02. Around 45\% and 32\% points exhibits weak and strong sub-ballistic behavior respectively. 
For \(\rho=0.5\), different regions of the contour plot in Fig.\ref{fig:6b} are colored in bright yellow or dark yellow i.e., all points exhibit strong sub-ballistic spreadings. 
Similarly, for \(\rho=0.8\), all the points in Fig. \ref{fig:6c} exhibits strong sub-ballistic spreading. The variation of the average entropy are shown in Figs.\ref{fig:6d}, \ref{fig:6e} and \ref{fig:6f} respectively for \(\rho=0.2, 0.5\mbox{ and }0.8\). The system exhibits \(\langle S_{E} \rangle \) values in the range \(0.59\leq \langle S_{E} \rangle < 0.99\) for \(\rho=0.2\). The dark blue colored elongated regions on two opposites sides of the contour plot shown in Fig.\ref{fig:6d} indicates the regions of lower \(\langle S_{E} \rangle \). The value of \(\langle S_{E} \rangle \) in those regions gradually increase with \(\rho\) as can be seen by comparing the related three plots for three different values of \(\rho\). The range of \(\langle S_{E} \rangle \) values changes to \(0.76\leq \langle S_{E} \rangle < 1\) and \(0.90\leq \langle S_{E} \rangle < 1\) respectively for \(\rho=0.5\) and 0.8. For \(\rho=0.5\),  nearly 38\% points exhibit \(\langle S_{E} \rangle \) values between 0.95 to 1. These points are situated within the bright yellow colored regions of the plot in Fig.\ref{fig:6e}. Lower entropy is again found in two elongated regions on two opposite sides of the contour plot. Nearly 8\% points exhibit \(\langle S_{E} \rangle \) values in the range \(0.99\leq \langle S_{E} \rangle < 1\) for \(\rho= 0.8\). These points are possible candidates for exhibiting maximal entanglement.\\

\subsection{Static Rudin-Shapiro Walk :\label{s3}} 

Static Rudin-Shapiro QW exhibits a spreading behavior which is the least sensitive to the coin parameter variation among the three different static QWs studied here. It predominantly exhibits a strong localized behavior. Figs. \ref{fig:7a}, \ref{fig:7b} and \ref{fig:7c} show the variations of \(\alpha\) for \(\rho=0.2,0.5\mbox{ and }0.8\) respectively. Major parts of all three contour plots are colored in dark blue indicating localized behavior. Nearly 95\%, 80\% and 57\% points exhibit \(\alpha < 0.25\) respectively for \(\rho=0.2,0.5 \mbox{ and }0.8 \). For all the three different values of \(\rho\), there are some points which exhibit negative values of \(\alpha\) indicating strong localization. It can be seen from Fig. \ref{fig:7a} that higher values of \(\alpha\) are exhibited mainly at two small elongated regions placed near the corners and oriented along the diagonal \(\theta=\phi\). The sizes of these two regions increase with \(\rho\). Nearly 0.43\%, 1.35\% \& 4.08\% points exhibit \(\alpha\) values in the range \(0.75\leq \alpha < 1\) for \(\rho=0.2, 0.5\mbox{ and }0.8\) respectively. These are the points which fall within the small regions. For \(\rho=0.5\), 4\% points exhibit \(\alpha\) values in the range  \(0.5\leq \alpha < 0.75\). The percentage increases to 15\% for \(\rho=0.8\). The variation of the average entropy are shown in Figs.\ref{fig:7d}, \ref{fig:7e} and \ref{fig:7f} respectively for \(\rho=0.2, 0.5\mbox{ and }0.8\). The system exhibits \(\langle S_{E} \rangle \) values in the range \(0.56\leq \langle S_{E} \rangle < 0.98\) for \(\rho=0.2\). The range of \(\langle S_{E} \rangle \) values changes to \(0.72\leq \langle S_{E} \rangle \leq .98\) and \(0.86\leq \langle S_{E} \rangle < .99\) respectively for \(\rho=0.5\) and 0.8. For \(\rho=0.2\), around 70\% points between 0.56 to 0.75. These points form the dark blue colored and bluish green colored regions shown in Fig.\ref{fig:7d}. Highest values of entropy are found in the bright yellow colored patch at the top right corner. With increasing values of \(\rho\), \(\langle S_{E} \rangle \) increases at different points. For \(\rho=0.5\), the dark blue region becomes more greenish in color. We find that nearly 53\% points \(0.70\leq \langle S_{E} \rangle \leq .80\) whereas only 8\% points exhibit entropy values in the range \(0.80\leq \langle S_{E} \rangle \leq .90\). The system exhibits highly entangled behavior for \(\rho=0.8\) as 24\% \& 55\% points exhibit entropy values in the ranges \(0.95\leq \langle S_{E} \rangle \leq .99\) and \(0.90\leq \langle S_{E} \rangle \leq .95\) respectively.\\

\subsection{Maximal Entanglement Generation :\label{s4}} The interesting phenomenon of maximal entanglement generation in quantum walk has been observed in case of binary dynamic disordered quantum walk whereas the static disordered walk does not exhibit such phenomenon \cite{maximal}. It is natural to ask whether time and position dependent arrangement of two coins based on deterministically aperiodic sequences can exhibit maximal entanglement. Liu et al. have recently showed that the dynamic and static Fibonacci walk exhibit maximal entanglement \cite{aperiodic_entanglement}. The result is quite interesting as maximal entanglement has not been found for static disordered QW. They studied the entanglement generation only for the following set of values of the coin parameters : (\(\rho_{1}=0.25,\rho_{2}=0.75,\theta=0,\phi=0\)) and (\(\rho_{1}=0.5,0\leq \rho_{2}\leq 1,\theta=0,\phi=0\)). They also took average over different parts of the sequence. On the contrary, we numerically generate a single long sequence whose length is larger than both the chain length and
the number of random walk steps. Moreover, we study the influence of coin parameters on the entanglement generation and the parameter values used here are quite different from that used by Liu et al.. Here, we have already seen that the dynamic Fibonacci walk, dynamic Rudin-Shapiro walk, the dynamic Thue-Morse walk, the static Fibonacci walk and the static Thue-Morse walk exhibits entropy values in the range \(0.99 \leq \langle S_{E}\rangle \leq 1\) at multiple values of \(\rho,\theta,\phi\). So, there is a high probability that many of those points exhibit maximal entanglement. Figs. \ref{fig:8a}, \ref{fig:8b}, \ref{fig:8c}, \ref{fig:8d} and \ref{fig:8e} show the variations of entropy \(S_{E}\) against time for the dynamic Fibonacci walk at (\(\rho = 0.8,\theta = 0.03  \pi, \phi = 1.46 \pi \)), the dynamic Rudin-Shapiro walk at (\(\rho = 0.8,\theta = 0 , \phi = 0.08 \pi \)), the dynamic Thue-Morse walk at (\(\rho = 0.8,\theta = 1.06  \pi , \phi = 0.80 \pi \)), the static Fibonacci walk at (\(\rho = 0.8,\theta = 1.06  \pi , \phi = 0.80 \pi \)) and the static Thue-Morse walk at (\(\rho = 0.8,\theta = 0.90  \pi , \phi = 1.41 \pi \)) respectively. These plots clearly indicate that the dynamic Fibonacci walk, the dynamic Rudin-Shapiro walk, the dynamic Thue-Morse walk can exhibit maximal entanglement generation at certain values of coin parameters. Similarly, the Static Fibonacci and static Thue-Morse walk also exhibited maximal entanglement generation in Fig.\ref{fig:8d} and \ref{fig:8e} respectively. We have shown here the plots of entropy variation against time only for a single (\(\theta,\phi\)) point. Maximal entanglement can also be found at many other points. However, we have skipped testing these large number of points individually as our interest is just in showing the capability of maximal entanglement generation of the binary deterministic aperiodic walks within the parameter space explored here. Thus all the walks studied here are capable of generating maximal entanglement under suitable choice of coin parameters except the static Rudin-Shapiro walk.\\  
\FloatBarrier
\section{Conclusion \& Outlook\label{four}}

The present work shows that the binary aperiodic quantum random walk phenomenon can be greatly influenced by the choice of quantum coins. It also shows that different aperiodic walks considered here respond quite differently to the coin parameter variations. The degrees of responses are found to be dependent on both the nature of the sequences and on the considered type of aperiodicity (dynamic or static). For example, the dynamic Fibonacci QW exhibits widest range of values of the exponent \(\alpha\) at \(\rho=0.2\). On the contrary, the dynamic Thue-Morse walk and the static Rudin-Shapiro walk exhibit much weaker dependence of \(\alpha\) on \(\theta,\phi\) at the same value of \(\rho\).\\

The present work also shows that aperiodic QWs can sometimes generate spreading behavior which is uncommon to both the periodic and disordered QWs. 
For example, the dynamic Fibonacci walk exhibits localized behavior for certain coin parameters. This is first time that a localization phenomenon is found for any type of QW with dynamic disorder/aperiodicity.\\


Previous works have shown that dynamic aperiodic walks spreads faster than the static aperiodic walks. Here, we demonstrate that certain coin parameters can reverse the scenario. The localization phenomenon has not been observed in static Fibonacci walk. This clearly indicates that for certain coin parameters, static Fibonacci walk spreads faster than dynamic Fibonacci walk.\\

The present work demonstrates for the first time the importance of the \(\rho\) parameter, which controls the superposition of the spin states, in controlling the (\(\theta,\phi\)) dependent spreading of a binary QW. For \(\rho=0.2\), \(\alpha\) and \(\langle S_{E}\rangle\) exhibits a broader range of values in most cases. On the contrary, for \(\rho=0.8\), \(\alpha\) and \(\langle S_{E}\rangle \) exhibits a shorter range of values. It indicates that \(\rho\) controls the sensitivity of \(\alpha\) and \(\langle S_{E}\rangle\) against variation of (\(\theta,\phi\)). For lower values of \(\rho\), the systems become more sensitive to \(\theta,\phi\) variations whereas the opposite happens for higher values of \(\rho\). We think that qualitatively same behavior might be found for binary periodic and randomly disordered walks. \\

Another motivation behind the present work was to check whether binary aperiodic sequences can exhibit similar behaver for certain coin conditions. Jing et al. found that sub-ballistic spreading could be found in binary dynamic disordered quantum walk if the two quantum coins satisfy the following condition :  \( e^{i(\theta_{1}-\phi_{1})}=e^{i(\theta_{2} -\phi_{2})}  \) where \(\theta_{1},\phi_{1},\theta_{2},\phi_{2}\) are coin parameters \cite{jing}. Similarly, the same condition was found to be required for binary static disordered walk to exhibit sub-diffusive, diffusive and sub-ballistic spreadings \cite{jing}. Here we find that binary aperiodic sequences do not follow such condition. The present work demonstrates that there is no such necessary/sufficient condition here for different spreading behavior. The aperiodic walks exhibit some similarity for some coin parameters. For example, high entanglement entropy is realized  at \(\rho=0.8\) for all different aperiodic QWs studied here, more or less independent of the other coin parameters. \\




Another interesting aspect of the present work is that this is the first time that entanglement is studied for full sequences of these aperiodic sequences. Moreover, we study the influence of generic coin parameters on the entanglement generation. Till now, the general idea is that the static disorder generates lower entropy in comparison to the dynamic disorder. We have found here a recipe to increase entropy for all these dynamic/static aperiodic sequences and it is to increase the value of the parameter \(\rho\). Apart from that, we have shown that Fibonacci and Thue-Morse QW, both generate maximal entanglement for both dynamic and static aperiodicity. The Rudin-Shapiro walk generates maximal entanglement only for dynamic aperiodicity.\\ 

One obvious extension of the present work would be to explore the influence of quantum coins on the other dynamical properties of the walker. It will be helpful in understanding the influence of quantum coins in more detail. Another future work would be to study coin influences for other initial states. We have considered here a localized initial state as done in most of the previous studies on binary aperiodic walks. However, similar studies with other initial states will help to answer the following question : How much does the behavior reported here depend on the initial state. The detailed nature of different anomalous spreadings reported here is also needed to be studied. One can also study the coin parameter variation for other aperiodic sequences. A thorough identification of the properties of any aperiodic QW based on coin parameter variation will be highly useful to successfully use it as a generator of various probability distributions and hybrid entanglements. The present work may be considered as one step towards understanding the role of coins in inhomogeniious quantum walks.\\

\bibliographystyle{revtex}
\bibliography{references}
{}

\end{document}